\documentclass[structabstract]{aa}  
\usepackage{graphicx}
\usepackage{txfonts}
\usepackage{epstopdf}

\begin{document}
   \title{An inhomogeneous lepto-hadronic model for the radiation of relativistic jets}

   \subtitle{Application to XTE J1118+480}

   \author{G. S. Vila,
          \inst{1}
   				G. E. Romero,\inst{1,2}
          \and
          N. A. Casco\inst{3}
%          \fnmsep\thanks{Just to show the usage
%          of the elements in the author field}
          }

   \institute{Instituto Argentino de Radioastronom\'ia (IAR - CONICET),
              C.C. 5 (1894), Villa Elisa, Buenos Aires, Argentina\\
              \email{gvila@iar.unlp.edu.ar}
         \and
             Facultad de Ciencias Astron\'omicas y Geof\'isicas, Universidad Nacional de La Plata (FCAG, UNLP), Paseo del Bosque s/n (1900), La Plata, Buenos Aires, Argentina\\
             \email{romero@iar.unlp.edu.ar}
             %\thanks{The university of heaven temporarily does not
              %       accept e-mails}
						\and Servicios Tecnol\'ogicos Integrados (STI), Lagos del Sur 63 (8400), San Carlos de Bariloche, R\'io Negro, Argentina\\
             \email{NCasco@sti-tech.com.ar}
             }
   \date{Received ; accepted }

% \abstract{}{}{}{}{} 
% 5 {} token are mandatory
 
  \abstract
   {Conceptually reconstructing the physical conditions in relativistic jets, given the observed electromagnetic spectrum, is a complex inverse problem. }
   { We aim to improve our understanding of the mechanisms operating in relativistic jets by modeling  their broadband electromagnetic spectrum.}
   {We develop an inhomogeneous jet model where the injection of relativistic primary and secondary particles takes place in a spatially extended region.  We calculate the contribution of all particles species to the jet emissivity by several radiative processes, and assess the effect of gamma-ray absorption in internal and external photon fields. A number of specific models with different parameters are computed to explore the possibilities of this scenario. }
   {We obtain a variety of spectral shapes depending on the model parameters, some of them predicting significant gamma-ray emission. The observed broadband spectrum of the low-mass microquasar XTE J1118+480 can be satisfactorily reproduced by the model.}
   {Our results indicate that outbursts similar to those displayed in the past by XTE J1118+480 might be detected with present-day gamma-ray instruments.}


\titlerunning{An inhomogeneous lepto-hadronic jet model}
\authorrunning{Vila et al.}
   \maketitle
% 
%________________________________________________________________

\section{Introduction}
\label{sec:introduction}

Jets are collimated outflows. Astrophysical objects that launch jets have two common components: a rotating compact object and an accretion disk. Examples of these systems are gamma-ray bursts, active galactic nuclei, microquasars, and young stellar objects.

Microquasars are X-ray binaries (XRBs) that produce relativistic jets. The jets of microquasars have bulk Lorentz factors of \mbox{$\sim1-10$}, and a typical power of \mbox{$10^{35-37}$ erg s$^{-1}$} (e.g. Gallo et al. 2005, Sell et al. 2010, see also Heinz \& Grimm 2005 and references therein).  Steady jets in microquasars appear in the so-called low-hard spectral state. In this state, the X-ray spectrum is the sum of a low-luminosity black body component and a hard power-law that cuts off at  \mbox{$\sim 100-150$} keV. These characteristics are well reproduced by ``disk+corona''  models (see for example Poutanen et al. 1997, Dove et al. 1997, and Malzac et al. 2001). In these models, the thermal emission originates in an accretion disk (truncated at tens or hundreds of gravitational radii from the compact object), and the hard power-law is generated in a ``corona'' of hot plasma that surrounds the black hole.  The main radiative mechanism of the corona is the Comptonization of disk photons by thermal electrons or electron-positron pairs.  The presence of the corona can also explain the detection of the Fe K$\alpha$ line at $\sim7$ keV in some XRBs (e.g. Miller et al. 2002, Tomsick et al. 2009, Duro et al. 2011). The transition from an outer thin disk to an inner quasi-spherical corona is self-consistently predicted in some accretion models, such as the advection-dominated accretion flows (ADAFs) studied by Narayan \& Yi (1994, 1995) and the adiabatic inflow--outflow solutions (ADIOS) of Blandford \& Begelman (1999). 

The ejection of jets  during the low-hard state is revealed by the flat radio spectrum, which is interpreted as synchrotron radiation of non-thermal electrons in an expanding outflow (Blandford \& K\"onigl 1979). Broadband observations of some XRBs indicate that there is a correlation of the form $S_{\rm{radio}}\propto S_X^{0.7}$ between the radio and the X-ray flux (Gallo et al. 2003). This suggests that the jets might contribute significantly to, or even dominate, the X-ray emission in the low-hard state. Two confirmed microquasar - Cygnus X-1 (McConnell et al. 2002, Albert et al. 2007, Sabatini et al. 2010, Laurent et al. 2011) and Cygnus X-3 (Abdo et al. 2009, Tavani et al. 2009) - have been detected in gamma rays. The jets - loaded with high-energy particles and moving at relativistic bulk velocities - appear as the probable site of the gamma-ray emission.\footnote{If a fraction of the particles in the corona are relativistic, the corona might also radiate gamma rays; see Romero et al. (2010) and Vieyro \& Romero (in preparation).} 

Jet radiative models have been typically divided into ``leptonic'' (e.g. Markoff et al. 2001, Kaufman Bernad\'o et al. 2002, Bosch-Ramon et al. 2006) and ``hadronic'' (e.g. Romero et al. 2003, 2005). In  hadronic (leptonic) models, the bulk of the gamma-ray emission is produced in interactions initiated by relativistic protons (electrons). Although microquasar jets can propagate over distances of hundreds of AU (Mirabel \& Rodr\'iguez 1999), most of these  models are based on the ``one-zone'' approximation: the emission region is assumed to be homogeneous and compact. 

In this work, we present an inhomogeneous lepto-hadronic model for the radiation of jets. We intend to expand available models by adopting more general assumptions about the contents of relativistic particles and the characteristics of the acceleration/emission region. In our model, relativistic protons and electrons are injected in a spatially extended and inhomogeneous region of the jet. We investigate the consequences of the propagation and cooling of these particles (and also of secondary muons, pions, and pairs) on the jet radiative output. The characteristics of the spectral energy distributions (SEDs) can provide insight into the physical conditions in the jets. As an application, we fit the spectrum of the low-mass XRB XTE J1118+480, a very well-studied black hole candidate in the Galactic halo. 

The paper is organized as follows. In Sect. \ref{sec:jet_model}, we introduce the main features of the model. We present the general assumptions made about the geometry and energetics of the jet, and briefly describe the interaction processes between the relativistic particles and the matter, radiation, and magnetic field. We outline the method of calculation of the particle distributions and the jet radiative spectrum, including absorption corrections. In Sect. \ref{sec:general_results}, we exhibit and analyze the results of a series of models calculated for different sets of parameters. Section \ref{sec:XTE} presents the application to XTE J1118+480. We show the best-fit SEDs and discuss the possible detection of the source with present and future gamma-ray telescopes. We close with a short review of the model results and the perspectives for future studies.

%__________________________________________________________________

\section{Inhomogeneous jet model}
\label{sec:jet_model}

\subsection{The jet}
\label{subsec:jet_parameters}

We generalize the one-zone model presented in Romero \& Vila (2008), Reynoso \& Romero (2009), and Vila \& Romero (2010) to treat the injection and cooling of relativistic particles in an extended inhomogeneous region of the jet.  

\begin{figure}%
\centering
\includegraphics[width=0.7\columnwidth, keepaspectratio, trim= 20 160 80 180, clip]{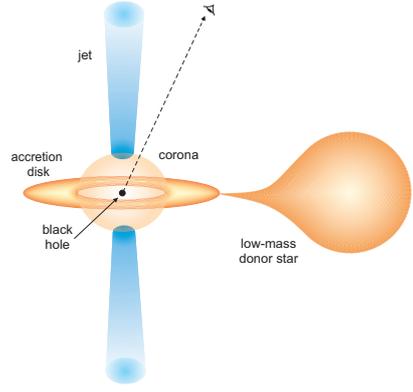}%
\caption{Sketch of a low-mass microquasar.}%
\label{fig:binary}%
\end{figure}

\begin{figure}%
\centering
\includegraphics[width=0.7\columnwidth, keepaspectratio, trim= 0 90 350 50, clip]{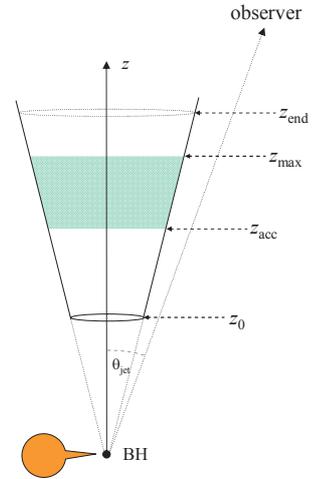}%
\caption{Detail of the jet and the acceleration region (not to scale). Some relevant parameters are indicated.}%
\label{fig:jet_detail}%
\end{figure}

The main components of the binary system are sketched in Fig. \ref{fig:binary}; a sketch of the jet is shown in Fig. \ref{fig:jet_detail}. A compact object (hereafter a black hole) of mass $M_{\rm{BH}}$ accretes matter from a low-mass companion star that has filled its Roche lobe.\footnote{In low-mass microquasars, the donor star is old and dim, and the stellar wind and radiation field are negligible targets for the relativistic particles in the jet.} We write the accretion power $L_{\rm{accr}}$ in terms of the Eddington luminosity of the black hole as

\begin{equation}
L_{\rm{accr}}\equiv \dot{M}c^2 = q_{\rm{accr}}\,L_{\rm{Edd}}\approx 1.3\times10^{38} q_{\rm{accr}}\left(\frac{M_{\rm{BH}}}{M_{\odot}}\right) \mathrm{erg  \,s}^{-1},
\label{eq:accretion_power}
\end{equation}

\vspace{0.2cm} 

\noindent where $\dot{M}$ is the mass accretion rate, $M_{\odot}$ is the solar mass, and $q_{\rm{accr}}$ is an adimensional parameter. An accretion disk extends from an inner radius $R_{\rm{in}}$ to an outer radius $R_{\rm{out}}$. Nearer the black hole, the plasma ``inflates'' and forms a hot, optically thin, corona.

A fraction of the accreted matter is ejected into two symmetrical jets, each carrying a power

\begin{equation}
L_{\rm{jet}}= \frac{1}{2}q_{\rm{jet}}\,L_{\rm{accr}},
\label{eq:jet_power}
\end{equation}

\vspace{0.2cm} 

\noindent where $q_{\rm{jet}}<1$. The jets are launched perpendicularly to the plane of the accretion disk at a distance $z_0$ from the black hole, and propagate up to $z=z_{\rm{end}}$. The jet axis makes an angle $\theta_{\rm{jet}}$ with the line of sight of an observer on Earth. The outflow expands as a cone as it advances, hence its radius grows as 

\begin{equation}
r(z)=r_0\left(\frac{z}{z_0}\right).
\label{eq:jet_radius}
\end{equation}

\noindent None of the parameters of the model depend on the radial coordinate (radial symmetry assumption). 

If the plasma is ejected by some magnetocentrifugal mechanism, the magnetic energy density at the base of the jet must be enough to set the plasma in motion.  We fix the value $B_0$ of the magnetic field at $z_0$ by equipartition between the magnetic and the kinetic energy densities

\begin{equation}
U_{\rm{mag}}(z_{\rm{0}}) =  U_{\rm{kin}}(z_{\rm{0}}).
\label{eq:equipartition_z_0}
\end{equation}

\vspace{0.2cm} 

\noindent The magnetic energy density decreases as the jet expands. We parameterize the dependence on $z$ of the magnetic field strength as

\begin{equation}
B(z)= B_0\left(\frac{z_0}{z}\right)^m,
\label{eq:magnetic_field}
\end{equation}

\vspace{0.2cm} 
 
\noindent where $m \geq 1$ (e.g. Krolik 1999). 

In the usually accepted model of jet acceleration (see for example Spruit 2010), a fraction of the magnetic energy is converted into kinetic energy of the plasma. The bulk Lorentz factor of the jet, $\Gamma_{\rm{jet}}$, then increases up to a terminal value. The behavior of  $\Gamma_{\rm{jet}}$ with the distance to the black hole can be studied both analytically and numerically using the equations of the MHD (e.g. Lyubarsky 2009, Tchekhovskoy et al. 2008, 2010); a simpler approach is presented in Reynoso et al. (2011). Here we simply adopt a constant value of $\Gamma_{\rm{jet}}$ for all $z$, leaving dynamical considerations for forthcoming works.

In the region  \mbox{$z_{\rm{acc}}\leq z \leq z_{\rm{max}}$}, some fraction of the jet power is transformed into kinetic energy of relativistic electrons and protons. We assume that the acceleration mechanism is particle diffusion through shock fronts, which is also known as diffusive shock acceleration. This assumption constrains the value of $z_{\rm{acc}}$, since for shock waves to develop the magnetic energy density of the plasma must be in sub-equipartition with the kinetic energy density (Komissarov et al. 2007). We then fix $z_{\rm{acc}}$ from the condition

\begin{equation}
U_{\rm{mag}}(z_{\rm{acc}})= \rho U_{\rm{kin}}(z_{\rm{acc}}),
\label{eq:z_acc_value}
\end{equation}

\vspace{0.2cm} 
 
\noindent where $\rho < 1$. The power transferred to the relativistic particles is 

\begin{equation}
L_{\rm{rel}}= q_{\rm{rel}}L_{\rm{jet}},
\label{eq:power_rel_particles}
\end{equation}

\noindent where $q_{\rm{jet}}<1$. The value of  $L_{\rm{rel}}$ is the sum of the power injected in both protons and electrons 

\begin{equation}
L_{\rm{rel}} = L_p + L_e.
\label{eq:power_rel_p_e}
\end{equation}

\vspace{0.2cm} 

\noindent In addition, we assume that $L_p$ and $L_e$ are simply related as

\begin{equation}
L_p  = a\,L_e.
\label{eq:parameter_a}
\end{equation}

\vspace{0.2cm}

\noindent If $a>1$ the jet is proton-dominated, whereas for $a<1$ most of the power is injected in relativistic electrons. 

\subsection{Particle cooling and maximum energy}
\label{sec:cooling_interactions}

The relativistic particles in the jet lose energy by radiation and as they exert adiabatic work on the walls of the expanding plasma. Several processes contribute to the radiative cooling, since particles can interact with the magnetic, radiation, and matter fields of the jet.

The most important interaction channel with the jet magnetic field is synchrotron radiation. For a charged particle of energy $E$ and mass $m$, the synchrotron energy-loss rate is (e.g. Blumenthal \& Gould 1970)

\begin{equation}
\left.\frac{dE}{dt}\right|_{\rm{synchr}} = -  \frac{4}{3}\left(\frac{m_e}{m}\right)^3\frac{c\sigma_{\rm{T}}\, U_{\rm{mag}}}{m_ec^2}\frac{E^2}{mc^2},
\label{eq:loss_rate_synchr}
\end{equation}

\vspace{0.2cm} 

\noindent where $c$ is the speed of light, $m_e$ is the electron mass, $\sigma_{\rm{T}}$ is the Thomson cross-section, and $U_{\rm{mag}}$ is the magnetic energy density. The ratio $\left(m_e/m\right)^3$ in Eq. (\ref{eq:loss_rate_synchr}) makes synchrotron cooling particularly efficient for the lightest particles.
 
Relativistic electrons also cool through inverse Compton (IC) scattering off the jet radiation field. We calculate the energy-loss rate in the Klein-Nishina regime following Blumenthal \& Gould (1970). We only consider the synchrotron field of primary electrons as a target field. To estimate the  density of synchrotron photons, we apply the ``local approximation'' of Ghisellini et al. (1985)

\begin{equation}
n_{\rm{synchr}}(\epsilon,z)\approx\varepsilon_{\rm{synchr}}(\epsilon,z)\,\frac{r_{\rm{jet}}(z)}{c\epsilon},
	\label{eq:sy_local}
\end{equation}  

\noindent where $\varepsilon_{\rm{synchr}}(\epsilon,z)$ is the synchrotron power per unit volume per unit energy at energy $\epsilon$.  We do not consider the IC cooling of protons, since the energy-loss rate for this process is much lower than that of other channels of interactions between energetic hadrons with radiation.

The interaction of relativistic protons with radiation can create electron-positron pairs (``photopair'' production)  

\begin{equation}
p+\gamma\rightarrow p+e^-+e^+.
	\label{photopair}
\end{equation}  

\noindent The photon threshold energy is $\sim1$ MeV in the rest frame of the proton. At higher energies, proton-photon collisions can also yield pions (``photomeson'' production). The two main channels are

\begin{equation}
p+\gamma\rightarrow p+a\pi^0+b\left(\pi^++\pi^-\right)
	\label{eq:photomeson1}
\end{equation}  

\noindent and

\begin{equation}
p+\gamma\rightarrow n+\pi^++a\pi^0+b\left(\pi^++\pi^-\right),
	\label{eq:photomeson2}
\end{equation}  

\noindent where the integers $a$ and $b$ are the pion multiplicities. Both channels have approximately the same cross-section, and a photon threshold energy of $\sim145$ MeV in the proton rest-frame. 

The subsequent decay of charged pions injects leptons and neutrinos
 
\begin{equation}
\pi^+\rightarrow\mu^++\nu_\mu, \quad \mu^+\rightarrow e^++\nu_{\rm{e}}+\overline{\nu}_\mu,
	\label{eq:piondecay1}	
\end{equation}

\begin{equation}
\pi^-\rightarrow\mu^-+\overline{\nu}_\mu, \quad \mu^-\rightarrow e^-+\overline{\nu}_{\rm{e}}+\nu_\mu,
	\label{eq:piondecay2}	
\end{equation}

\noindent whereas neutral pions decay into two photons,

\begin{equation}
\pi^0\rightarrow\gamma+\gamma.
	\label{eq:piondecay3}	
\end{equation}

For photomeson and photopair production, the proton energy-loss rate can be calculated as in Begelman et al. (1990). Once again, we only consider as a target radiation field the synchrotron photons of primary electrons calculated with Eq. (\ref{eq:sy_local}). 

High-energy protons can also interact with the thermal particles in the jet plasma. If the energy of the relativistic proton is higher than the threshold value for $\pi^0$ production ($\approx1.22\,$ GeV), the collision with a thermal proton can create pions

\begin{equation}
p+p\rightarrow p+p+a\pi^0+b\left(\pi^++\pi^-\right),
	\label{pp1}
\end{equation}  

\begin{equation}
p+p\rightarrow p+n+\pi^++a\pi^0+b\left(\pi^++\pi^-\right).
	\label{pp2}
\end{equation}    

\noindent As in proton-photon interactions, proton-proton collisions inject photons by means of the decay of neutral pions. Charged pions inject electron-positron pairs and neutrinos through the decay chains of Eqs. (\ref{eq:piondecay1}) and (\ref{eq:piondecay2}). 

The energy-loss rate for a proton of energy $E$ is given by (e.g. Begelman et al. 1990)

\begin{equation}
\left.\frac{dE}{dt}\right|_{pp} = -n_p\,c\,E\,\kappa_{pp}\sigma_{pp},
	\label{eq:pp_loss_rate}
\end{equation}    

\noindent where $n_p$ is the number density of thermal protons, $\sigma_{pp}$ is the proton-proton cross-section, and \mbox{$\kappa_{pp}\approx0.5$} is the total inelasticity of the interaction. A convenient parametrization for $\sigma_{pp}$ is presented in Kelner et al. (2006). We calculate the value of $n_p$ in the observer rest-frame\footnote{Henceforth, we refer to as ``comoving'' or ``jet frame''  the reference frame fixed to the jet plasma, where the outflow bulk velocity is zero. Its $z$-axis is parallel to the jet symmetry axis. The ``observer frame'' correspond to that where the jet bulk velocity is $v_{\rm{jet}}$. When it is necessary to distinguish between the same quantity in both frames, the value in the jet frame is denoted by primed symbols.}  as in Bosch-Ramon et al. (2006)

\begin{equation}
n_{p}=\left(1 - q_{\rm{jet}}\right)\frac{L_{\rm{jet}}}{\Gamma_{\rm{jet}}\pi r_{\rm{jet}}^2v_{\rm{jet}}m_pc^2},
	\label{eq:thermal_particles_density} 
\end{equation} 

\noindent where $m_p$ is the proton mass.

Leptons can cool efficiently by interacting with the thermal protons in the jet through relativistic Bremsstrahlung. The energy-loss rate for a particle of energy $E$, charge $e$, and mass $m$ is

\begin{equation}
\left.\frac{dE}{dt}\right|_{\rm{Bremsstr}} = -4\,r_e^3\,\alpha_{\rm{FS}}\,c\,n_p(z)\left[\log\left(\frac{2E}{mc^2}\right)	- \frac{1}{3}\right]\,E,
\label{eq:bremsstr_cooling_rate} 
\end{equation} 

\vspace{0.2cm}

\noindent where $r_e$ is the classical electron radius and $\alpha_{\rm{FS}}$ is the fine structure constant.

Finally, to calculate the adiabatic energy loss rate we use the expression given in Bosch-Ramon et al. (2006)

\begin{equation}
\left.\frac{dE}{dt}\right|_{\rm{ad}}(E,z) = -\frac{2}{3}\frac{v_{\rm{jet}}}{z}E.
\label{eq:adiab_loss_rate}
\end{equation}

\vspace{0.2cm} 

\noindent The maximum energy of primary electrons and protons at $z$ can be estimated by equating the total energy-loss rate to the acceleration rate

\begin{equation}
-\left.\frac{dE}{dt}\right|_{\rm{tot}}(E_{\rm{max}},z) = \left.\frac{dE}{dt}\right|_{\rm{acc}}(E_{\rm{max}},z). 
\label{eq:max_energy}
\end{equation}

\vspace{0.2cm} 

\noindent The total energy-loss rate is simply the sum of the relevant loss rates for the particle species in consideration

\begin{equation}
\left.\frac{dE}{dt}\right|_{\rm{tot}}(E,z) = \sum_i{\left.\frac{dE}{dt}\right|_{i}(E,z)}.
\label{eq:total_loss_rate}
\end{equation}

\vspace{0.2cm} 

\noindent The acceleration rate is (e.g. Aharonian 2004)

\begin{equation}
	\left.\frac{dE}{dt}\right|_{\rm{acc}} = \eta\,e\,c\,B(z),
	\label{eq:acc-rate}
\end{equation}

\noindent where $e$ is the electron charge, and  $\eta<1$ is a phenomenological parameter that characterizes the efficiency of the acceleration mechanism. 

Pions and muons can also interact before decaying. This effect becomes important when particles are very energetic and the physical conditions in the jet are such that the cooling time is shorter than the mean lifetime.  

Our treatment of the cooling of secondary particles includes adiabatic and synchrotron losses. For muons and electron-positron pairs, we also consider IC scattering, and pion-photon and pion-proton collisions for charged pions. The cross-section of pion-proton interactions is $\approx 2/3$ of that for proton-proton collisions (Gaisser 1990). We apply the same approximation for pion-photon collisions.

The injection of secondary particles is not a consequence of diffusive shock acceleration, but a by-product of the interaction of primary protons and electrons.  The maximum energy of pions, muons, and secondary electron-positron pairs is then not fixed by Eq. (\ref{eq:max_energy}). It is instead determined by the characteristics of the particular process through which they were injected, and by the maximum energy of the primary particles.

\subsection{The steady-state distribution of relativistic particles}

We parameterize the particle injection function of primary electrons and protons in the jet reference frame (in units of erg$^{-1}\,$ s$^{-1}\,$cm$^{-3}$)  as a power-law in energy multiplied by an exponential cutoff 

\begin{equation}
Q(E,z) = Q_0\,E^{-\alpha}\exp\left[-E/E_{\rm{max}}(z)\right]\,f(z).
\label{eq:injection_function}
\end{equation}

\vspace{0.2cm} 
 
\noindent For diffusive shock acceleration, the value of the spectral index is typically in the range $1.5\lesssim\alpha\lesssim2.4$ (Rieger et al. 2007). The cutoff energy $E_{\rm{max}}$ depends on $z$, and is calculated by balancing the energy-loss rate and the acceleration rate, as explained in Sect. \ref{sec:cooling_interactions}. The function $f(z)$ is a step-like function that tapers the length of the acceleration region

\begin{equation}
f(z) = 1 - \frac{1}{1 + \exp[-(z - z_{\rm{max}})]}\approx 
 \left\{ \begin{array}{rc}
                    1 & \;\; z\leq z_{\rm{max}}, \\[0.1cm]
                    0 & \;\; z > z_{\rm{max}}.
         \end{array}
\right.
\label{eq:step_function}
\end{equation}

\vspace{0.2cm} 

\noindent Finally, the value of the normalization constant $Q_0$ is obtained from the total power injected in each particle species

\begin{equation}
L_i = \int_V d\vec{r}\int_{E_{\rm{min}}}^{E_{\rm{max}}}\,dE\, E\,Q(E,z),
\label{eq:calculation_q0}
\end{equation}

\vspace{0.2cm} 

\noindent where the index $i$ runs over protons and electrons. 

As shown in Sect. \ref{sec:cooling_interactions}, the interaction of relativistic protons with matter and radiation injects charged pions, muons, and secondary electron-positron pairs. Electron-positron pairs are also created through photon-photon annihilation.  The expression of $Q(E,z)$ for secondary particles depends on the specific injection process. 

The sources of charged pions are proton-photon and proton-proton interactions. Suitable expressions for the corresponding pion injection functions were presented in  Atoyan \& Dermer (2003), Kelner et al. (2006), and Kelner \& Aharonian (2008). We use the formulas of Lipari et al. (2007) for the muon injection function. Finally, to calculate the electron-positron  injection rate from muon decay we follow Schlickeiser (2002).

To estimate the pair output from photopair production, we apply the formulas of Chodorowski et al. (1992) and M\"ucke et al. (2000). The last source of pair injection that we consider is photon-photon annihilation

\begin{equation}
\gamma+\gamma\rightarrow e^++e^-.
	\label{eq:gamma-gamma}
\end{equation}

\noindent The pair injection function for this process is derived in B\"ottcher \& Schlickeiser (1997). Two-photon annihilation is also a photon sink, and can eventually modify the production spectrum. Its effect on the escape of photons from the jet is discussed later.

Once the injection function is known, we calculate the isotropic, steady-state particle distributions $N(E,z)$ (in units of erg$^{-1}\,$cm$^{-3}$) in the jet reference frame by solving a transport equation of the form (e.g. Khangulyan et al. 2008) 

\begin{equation}
v_{\rm{conv}}\frac{\partial N}{\partial z} + \frac{\partial}{\partial E} \left(\left.\frac{dE}{dt}\right|_{\rm{tot}}N\right)+ \frac{N}{\tau_{\rm{dec}}(E)} = Q(E,z).
\label{eq:transport_equation}
\end{equation}

\vspace{0.2cm} 

\noindent This equation is appropriate to describe  particle distributions over spatially extended regions. It incorporates the effects of the variation with $z$ in the parameters that govern the interaction of the relativistic particles, namely the magnetic field, the radiation fields, and the density of thermal particles. From left to right, the three terms on the left-hand side of Eq.(\ref{eq:transport_equation}) account for the changes in $N(E,z)$ caused by particle convection along the jet, energy losses, and decay.  The convection velocity is on the order of the jet bulk velocity, $v_{\rm{conv}}\approx v_{\rm{jet}}$. The decay term is nonzero only for pions and muons, where $\tau_{\rm{dec}}(E)$ is the corresponding mean lifetime in the jet frame. 

\subsection{Non-thermal spectrum of the jet}

We calculate the radiative spectrum of all particle species produced by each of the interaction processes described in Sect. \ref{sec:cooling_interactions}. Detailed formulas can be found in Romero \& Vila (2008), Reynoso \& Romero (2009), and references therein. 
 
The spectra are first obtained in the co-moving frame, and then transformed to the observer frame. If $q_{\gamma}^{\prime\,(i)}$ is the jet-frame photon emissivity (in erg s$^{-1}$ cm$^{-3}$)  associated with process $i$,  the specific luminosity at energy $E'_\gamma$ is

\begin{equation}
	L_{\gamma}^{\prime\,(i)}(E'_\gamma) = 2\pi\int_{z_{\rm{acc}}}^{z_{\rm{max}}} r_{\rm{jet}}^2(z) \,q_{\gamma}^{\prime\,(i)}(E'_\gamma, z)\,dz.
	\label{eq:luminosity}
\end{equation}

\vspace{0.2cm}

\noindent The luminosity in the observer frame is then

\begin{equation}
L^{(i)}_\gamma(E_\gamma)=D^2\,L_\gamma^{\prime\,(i)}(E_\gamma^\prime),
	\label{eq:boost-luminosity}
\end{equation}

\noindent where 

\begin{equation}
E_\gamma=DE_\gamma^\prime
	\label{eq:boost-energy}
\end{equation}

\noindent is the photon energy in the observer frame, and

\begin{equation}
D=\left[\Gamma_{\rm{jet}}\left(1-\beta_{\rm{jet}}\cos\theta_{\rm{jet}}\right)\right]^{-1}
	\label{eq:boost-factor}
\end{equation}

\vspace{0.2cm}

\noindent is the Doppler factor. In Eq. (\ref{eq:boost-factor}), $\theta_{\rm{jet}}$ is the jet viewing angle and $\beta_{\rm{jet}}=v_{\rm{jet}}/c$. In the particular case of proton-proton interactions, the luminosity is calculated directly in the observer frame. Since Eq. (\ref{eq:transport_equation}) yields the proton distribution in the jet frame, it must be first transformed to the observer frame.  For this, we use the general expressions given in Torres \& Reimer (2011).

The emission spectrum must be corrected for absorption caused by photon-photon annihilation into electron-positron pairs (see Eq. (\ref{eq:gamma-gamma})). The optical depth for a photon of energy $E_\gamma$ created at height $\tilde{z}$ on the jet axis is 

\begin{eqnarray}
\tau_{\gamma\gamma}(E_{\gamma},\tilde{z})=\frac{1}{2}\int_{0}^{\infty}\,\,\int^{u_{\rm max}}_{-1}\int^{\epsilon_{\rm max}}_{\epsilon_{\rm th}} \,\sigma_{\gamma\gamma}(E_{\gamma},\epsilon,u)\,n(\epsilon,\vec{r})\,\times  \;\nonumber\\ (1-u)\,d\epsilon\;\,du\;dl,
	\label{eq:tau_gamma-gamma}
\end{eqnarray}

\noindent where $l$ is the length of the path traversed by the photon from the emission site to the observer, $n(\epsilon,\vec{r})$ is the energy distribution of targets photons, $u$ is the cosine of the collision angle, and $\sigma_{\gamma\gamma}$ is the annihilation cross-section (e.g. Gould \& Schr\'eder 1967). The threshold energy of the target photon depends on the collision angle, and is given by  

\begin{equation}
\epsilon_{\rm{th}}(E_\gamma,u) =\frac{2m_e^2c^4}{E_\gamma\left(1-u\right)}. 
	\label{eq:gamma-gamma-threshold}
\end{equation} 

\noindent  Eq. (\ref{eq:tau_gamma-gamma}) is simplified when the target photon distribution is isotropic. A convenient expression for $\tau_{\gamma\gamma}$ in this particular case is obtained in Gould \& Schr\'eder (1967).

%______________________________________________________________

\section{General results}
\label{sec:general_results}

We compute several models with different values of some of the parameters. Here we present the results for four representative models.  In all cases, we fix \mbox{$M_{\rm{BH}} = 10 M_\odot$}, \mbox{$z_0 = 50\,R_{\rm{grav}}$},\footnote{$R_{\rm{grav}}$ is the gravitational radius of the black hole, $R_{\rm{grav}}= GM_{\rm{BH}}/c^2$.} \mbox{$r_0=0.1\,z_0$}, \mbox{$z_{\rm{end}} = 10^{12}$ cm}, \mbox{$\theta_{\rm{jet}} = 30^\circ$}, \mbox{$q_{\rm accr} = 0.1$}, \mbox{$q_{\rm jet} = 0.1$}, \mbox{$q_{\rm rel} = 0.1$}, \mbox{$\eta = 0.1$}, and \mbox{$E_{\rm{min}} = 10\,m_{(e,p)}c^2$}. The parameters specific to each model are listed in Table \ref{tab:general_models}.

\begin{table*}[htb]
\begin{center}
\caption{Values of the model parameters for four representative models.}
\label{tab:general_models}
\begin{tabular}{@{}p{5cm}|c|c|c|c}
\hline
\hline
Parameter (symbol) & Model A & Model B & Model C & Model D \\
\hline
&&\\[-0.3cm] 
Magnetic field decay index $(m)$ 						      & 1.5 & 2.0 &  1.5 & 1.5 \\[0.15cm]
Ratio $U_B/U_{\rm{k}}$ at $z_{\rm{acc}}$ $(\rho)$ & 0.9 & 0.9 &  0.5 & 0.1 \\[0.15cm]
Base of the acceleration region $(z_{\rm acc})$ 	& $1.8\times10^{8}$ cm & $1.2\times10^{8}$ cm &  $3.2\times10^{8}$ cm & $1.6\times10^{9}$ cm\\[0.15cm]
End of the acceleration region $(z_{\rm max})$ 		& $10^{10}$ cm & $10^{10}$ cm &  $10^{11}$ cm & $10^{10}$ cm \\[0.15cm]
Ratio $L_p/L_e$  $(a)$ 														& 1 & 100 &  1 & 1 \\[0.15cm]
Power relativistic protons  $(L_p)$ 							& $3.2\times10^{35}$ erg s$^{-1}$ & $6.4\times10^{35}$ erg s$^{-1}$  & $3.2\times10^{35}$ erg s$^{-1}$ & $3.2\times10^{35}$ erg s$^{-1}$\\[0.1cm]
Power relativistic electrons $(L_e)$ 							& $3.2\times10^{35}$ erg s$^{-1}$ & $6.4\times10^{33}$ erg s$^{-1}$ & $3.2\times10^{35}$ erg s$^{-1}$ & $3.2\times10^{35}$ erg s$^{-1}$ \\[0.1cm]
Particle injection spectral index $(\alpha)$ 			& 1.5 & 1.5 &  1.5 & 2.2 \\[0.15cm]
\hline
\hline
\end{tabular}
\end{center}
\end{table*}

\subsection{Cooling rates}
\label{sub-sec:cooling_rates}

\begin{figure*}[htb]
\centering
\includegraphics[width=0.33\textwidth]{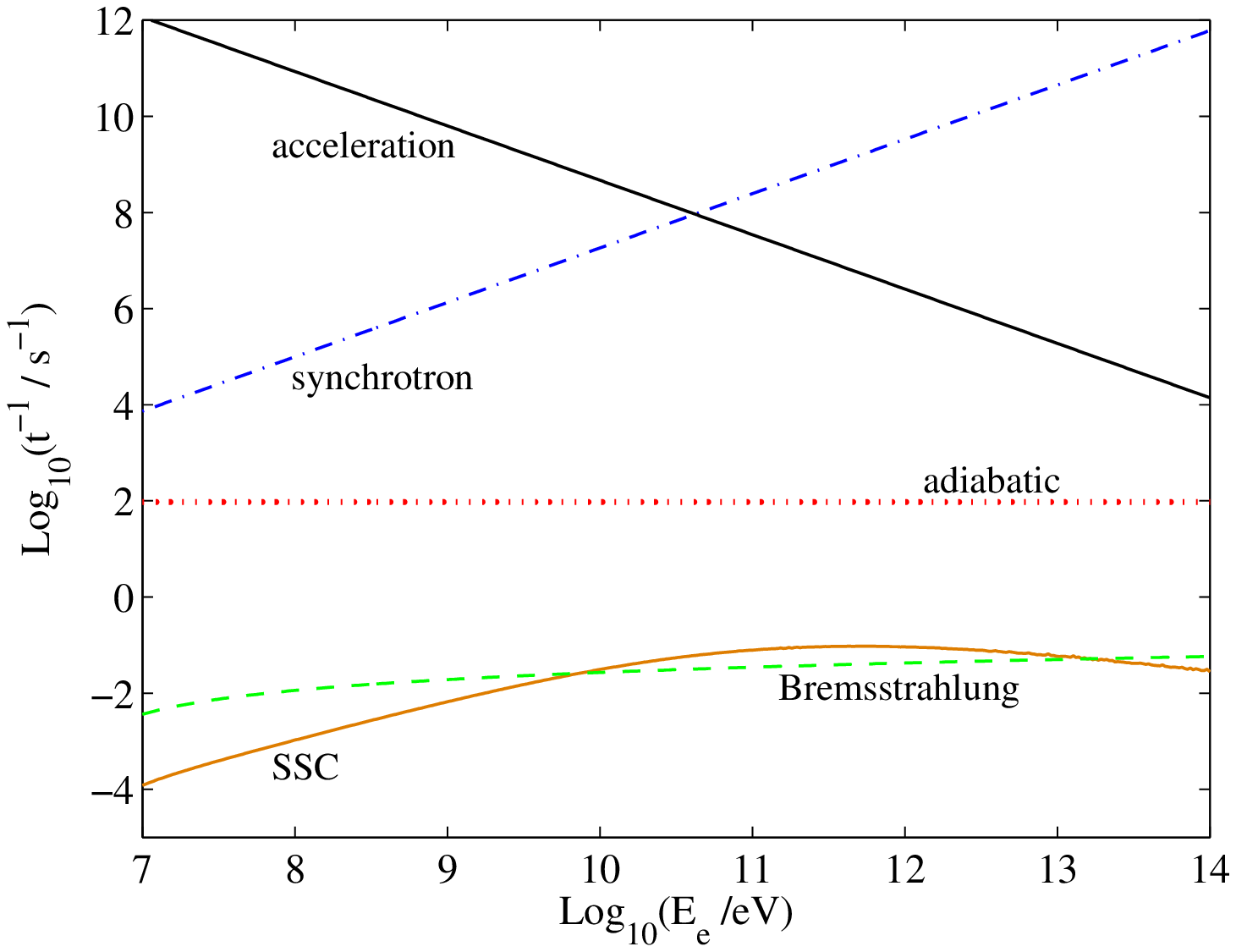}%
\includegraphics[width=0.33\textwidth]{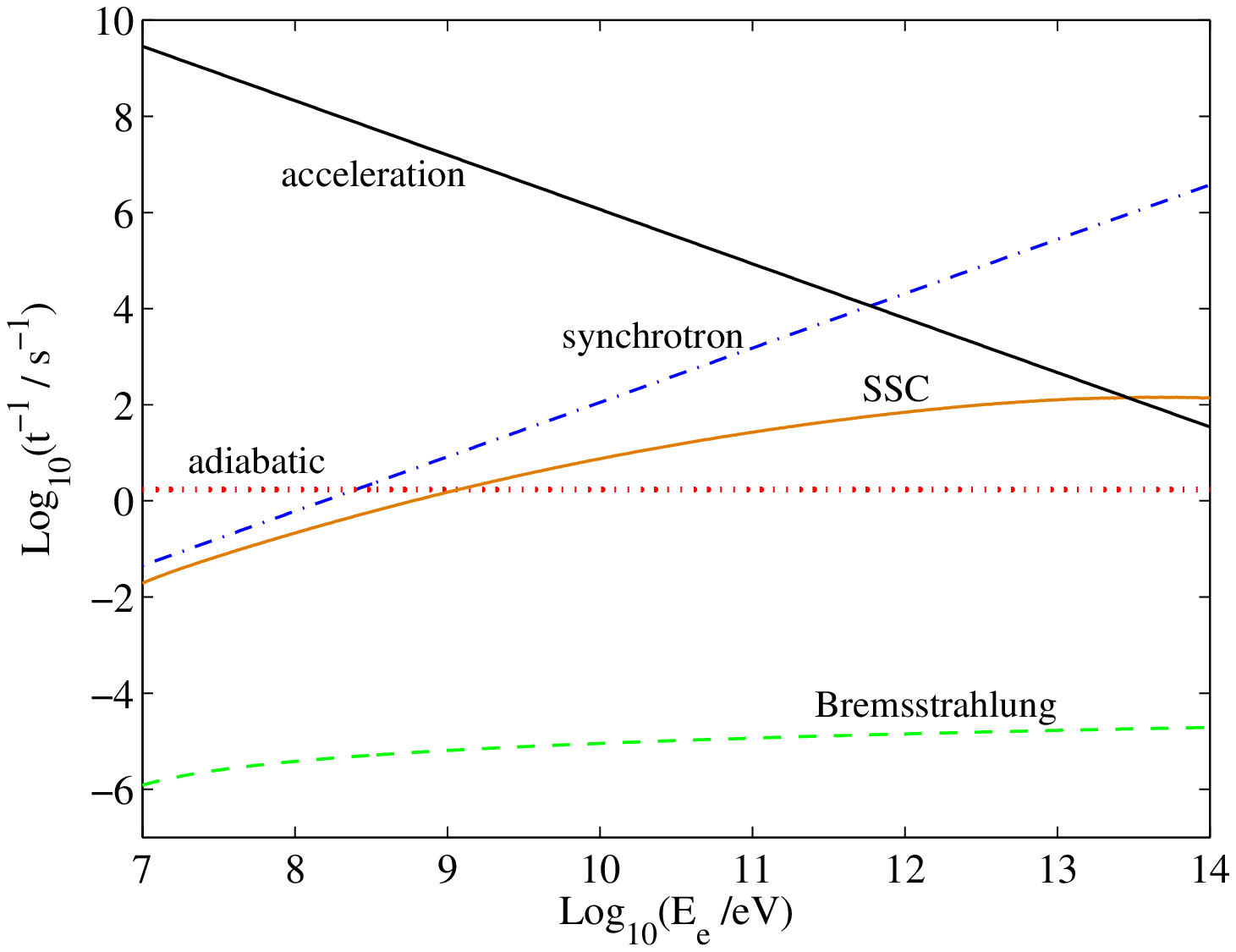}%
\includegraphics[width=0.33\textwidth]{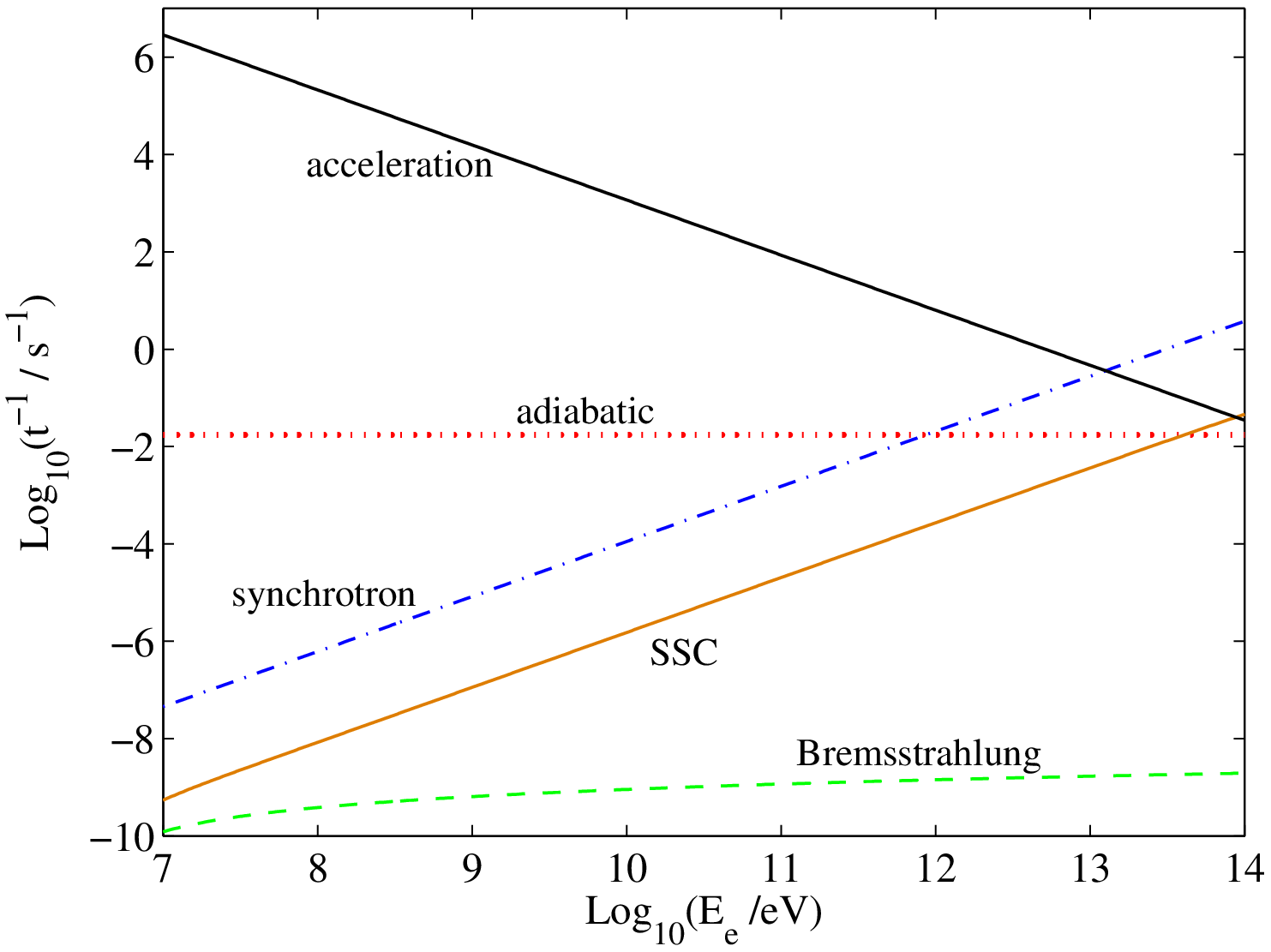}%
\caption{Cooling and acceleration rates for relativistic electrons for model A at the base (left) and the top (center) of the acceleration region, and at the ``end'' of the jet (right).}%
\label{fig:cooling_electrons}%
\end{figure*}

\begin{figure*}[htb]%
\centering
\includegraphics[width=0.33\textwidth]{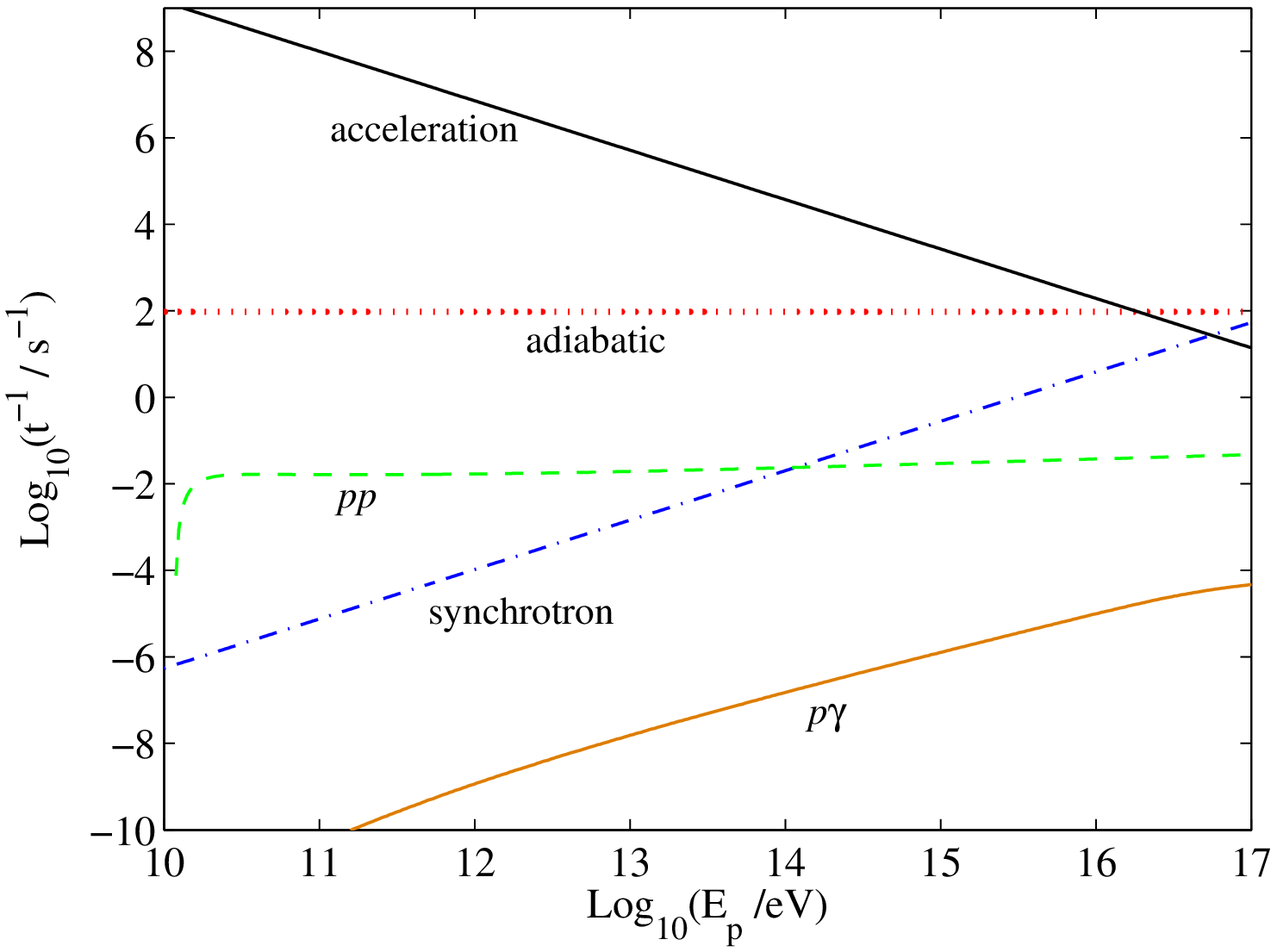}%
\includegraphics[width=0.33\textwidth]{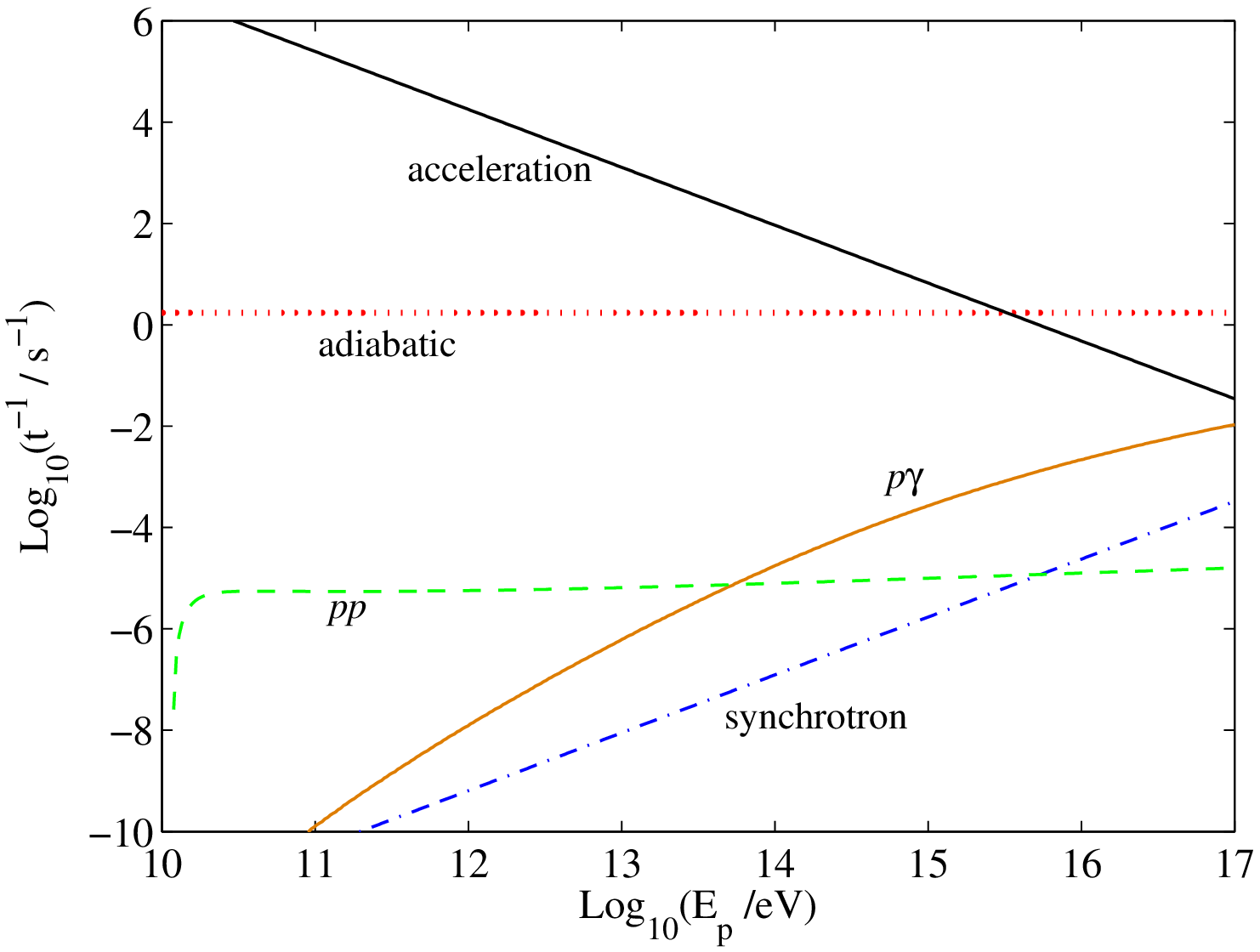}%
\includegraphics[width=0.33\textwidth]{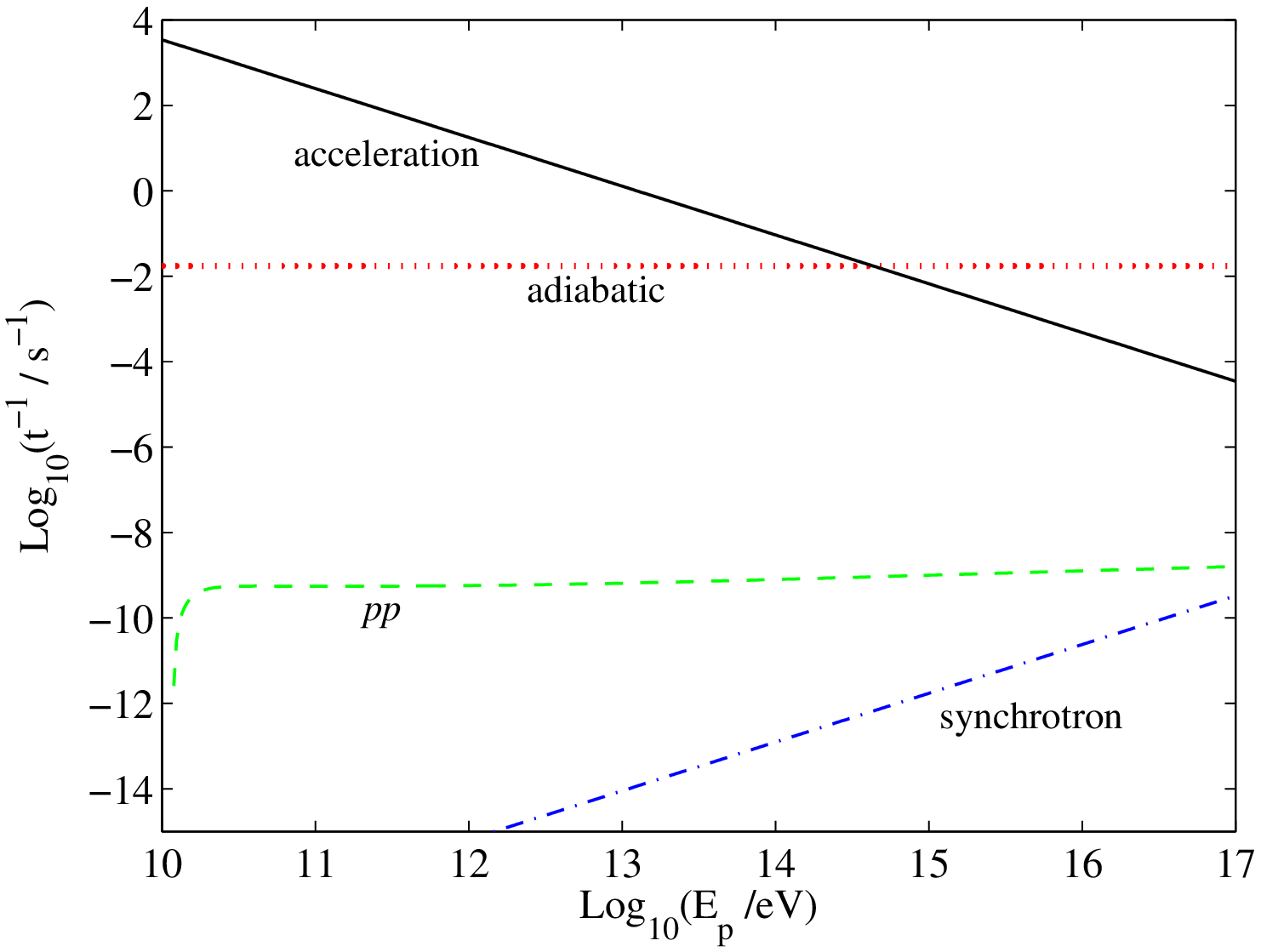}%
\caption{The same as in Fig. \ref{fig:cooling_electrons} but for relativistic protons. The abbreviations ``$pp$'' and ``$p\gamma$'' stand for proton-proton and proton-photon, respectively. }%
\label{fig:cooling_protons}%
\end{figure*}

In Figs. \ref{fig:cooling_electrons} and \ref{fig:cooling_protons}, we plot the cooling rates for primary electrons and protons in model A. They are calculated at $z=z_{\rm{acc}}$ (base of the acceleration region), $z=z_{\rm{max}}$ (top of the acceleration region), and $z=z_{\rm{end}}$ (``end'' of the jet). The cooling (acceleration) rate for process $i$ is defined as

\begin{equation}
	t^{-1}_{i} = -\frac{1}{E}\left.\frac{dE}{dt}\right|_{i}.
	\label{eq:cooling-rate}
\end{equation}

\noindent Synchrotron radiation dominates the cooling of electrons near the base of the acceleration region. Further away from the black hole, this process gradually becomes less relevant at lower energies, since $t^{-1}_{\rm{synchr}}\propto z^{-3}$. The maximum energy of electrons, however, is always determined by the balance of the synchrotron cooling rate and the acceleration rate. Synchrotron-self-Compton\footnote{Synchrotron-self-Compton (SSC) scattering  is the IC scattering of particles's own synchrotron radiation field.} (SSC) energy losses are in general much smaller, and the cooling caused by relativistic Bremsstrahlung is negligible. 

Adiabatic energy losses are the most important for protons all along the jet. We note that the cooling produced by proton-photon $(p\gamma)$ interactions is completely negligible at large $z$. 

\subsection{Particle injection and distributions}
\label{sub-sec:particle_distributions}

\begin{figure*}%
\centering
\includegraphics[width=0.49\textwidth, keepaspectratio]{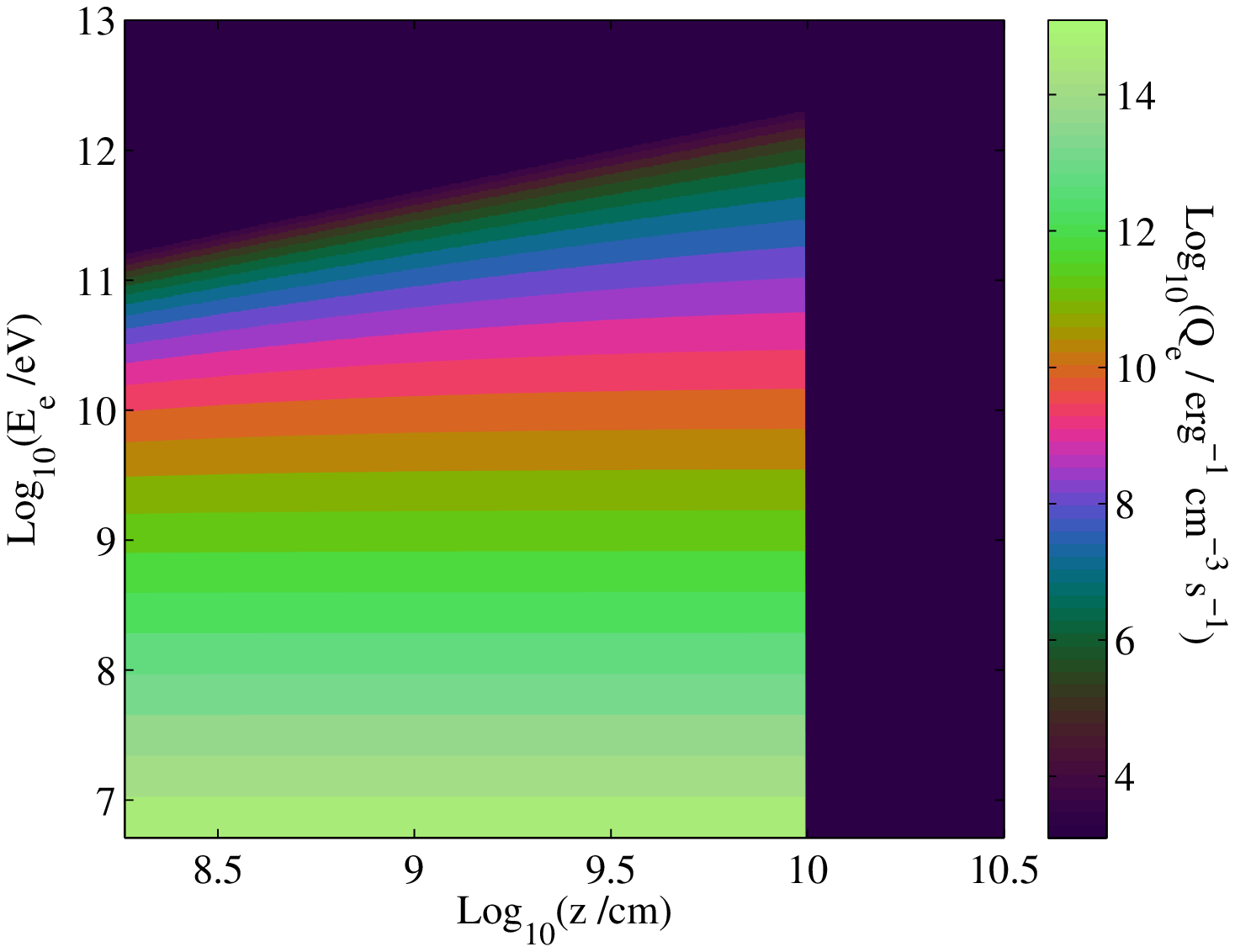}%
\includegraphics[width=0.49\textwidth, keepaspectratio]{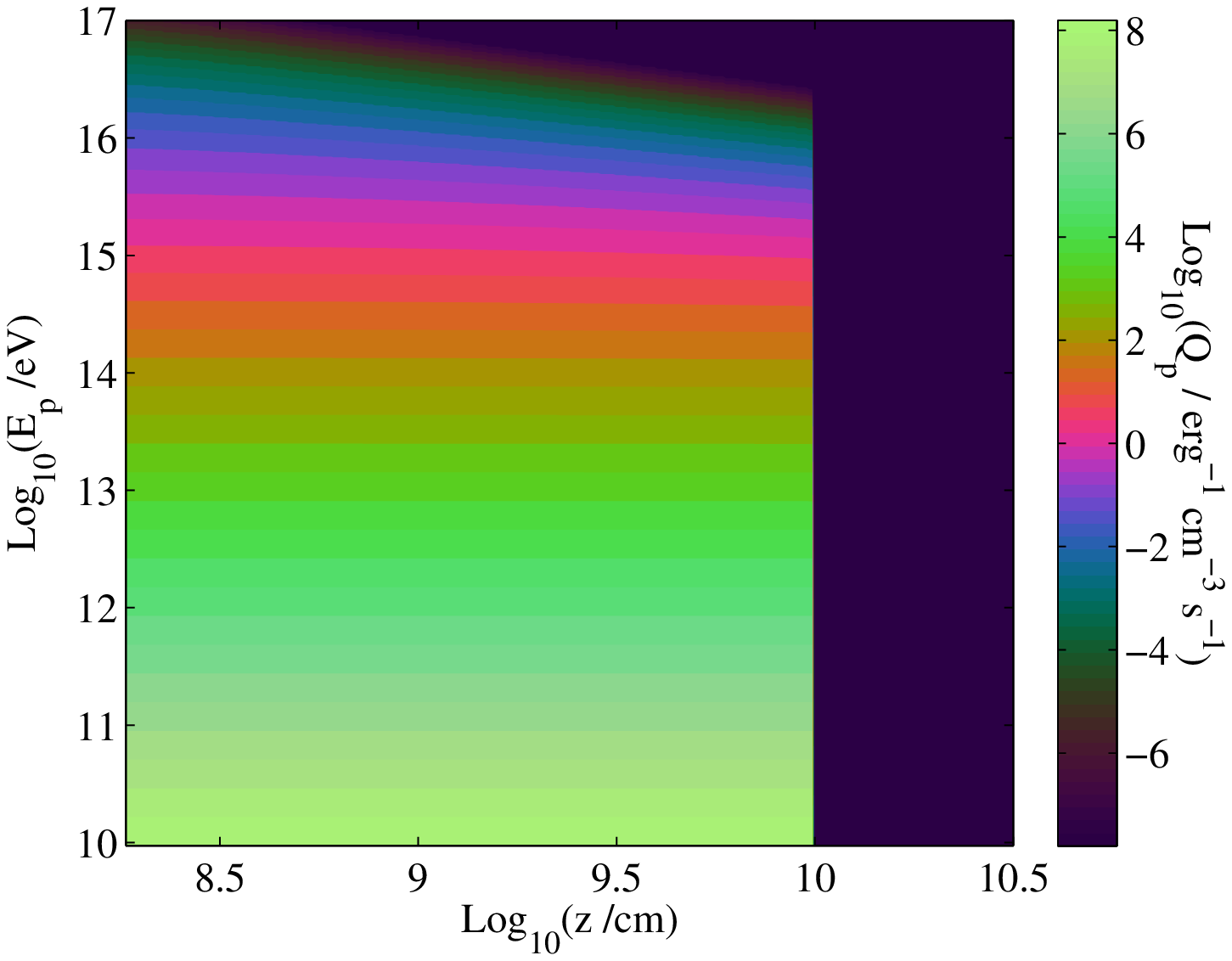}%
\caption{Injection function of relativistic electrons (left) and protons (right) for model A. Figures available in color in the electronic version of the manuscript.}%
\label{fig:injections}%
\end{figure*}

Figure \ref{fig:injections} shows the dependence of the injection function on energy and $z$, for primary electrons and protons in model A. The injection is confined to the region $z < z_{\rm{max}} = 10^{10}$ cm. As expected from the cooling rates in Figs. \ref{fig:cooling_electrons} and \ref{fig:cooling_protons}, protons reach energies much higher than electrons. The maximum energy of electrons is determined by the synchrotron losses, and hence it increases with $z$ as the magnetic field decreases. For protons, adiabatic losses are the main cooling channel, and the maximum proton energy decreases with $z$. 

\begin{figure*}%
\centering
\includegraphics[width=0.49\textwidth, keepaspectratio]{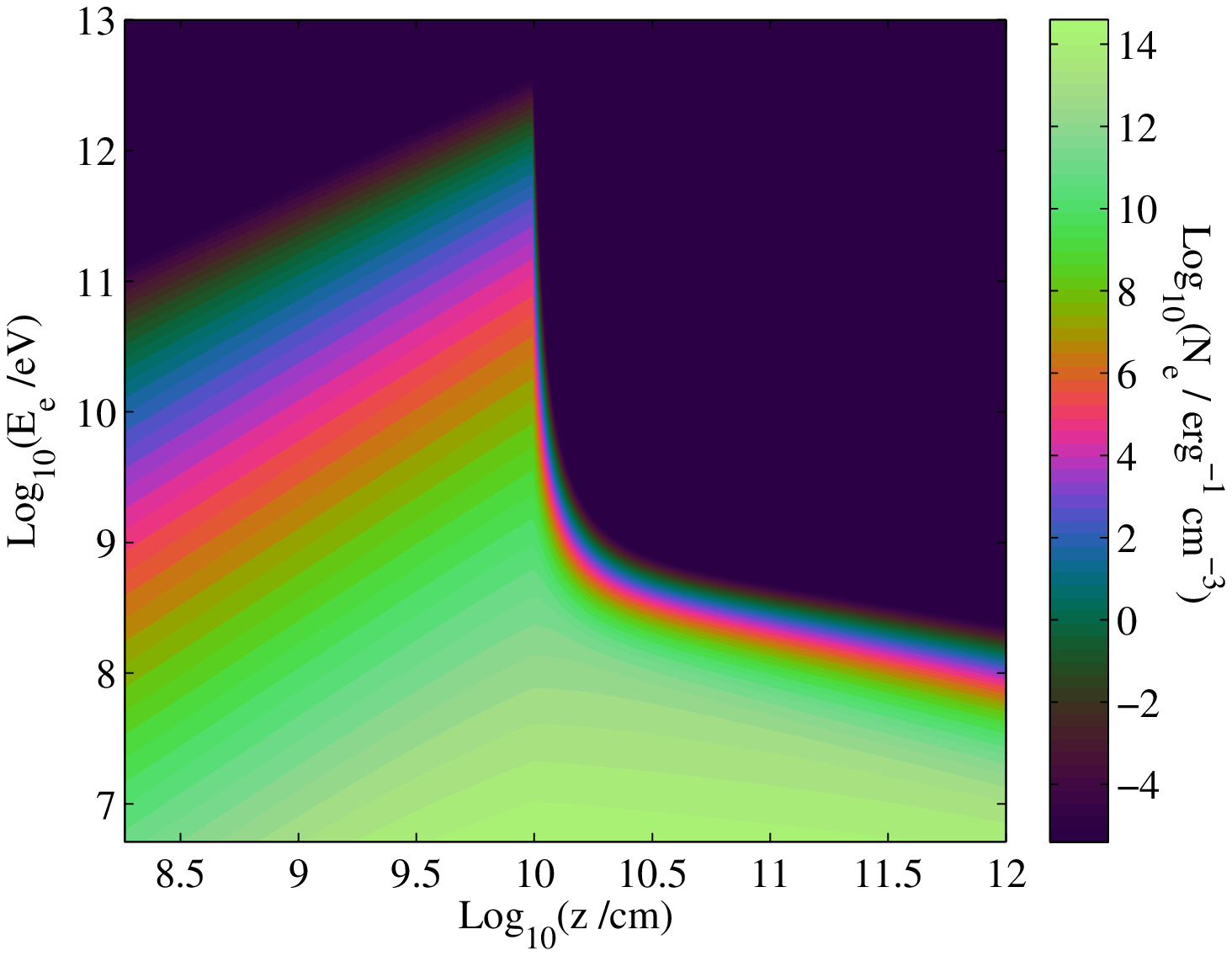}%
\includegraphics[width=0.49\textwidth, keepaspectratio]{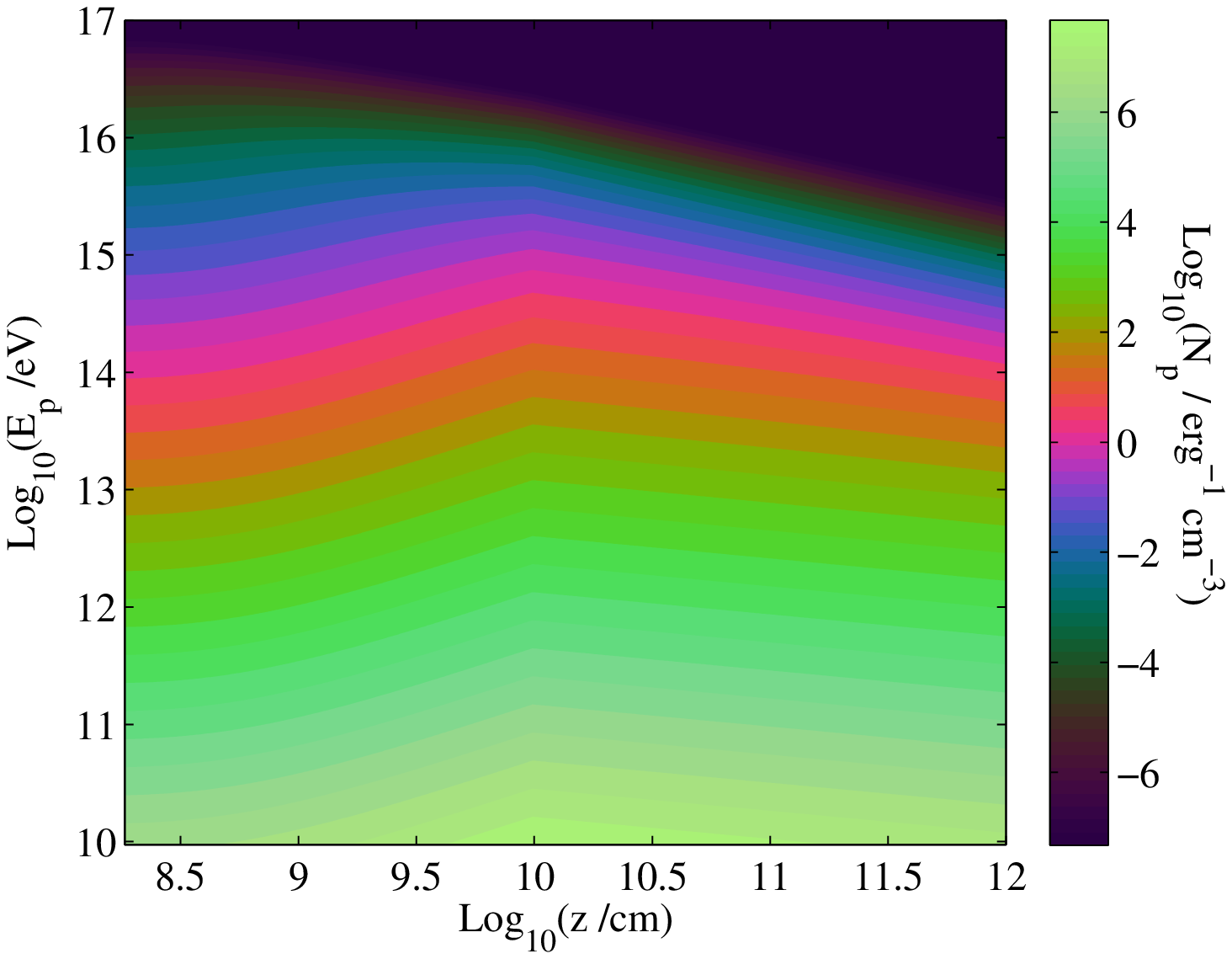}%
\caption{Energy distribution of relativistic electrons (left) and protons (right) for model A. Figures available in color in the electronic version of the manuscript.}%
\label{fig:distributions}%
\end{figure*}

The steady-state particle distributions calculated from Eq. (\ref{eq:transport_equation}) are plotted in Fig. \ref{fig:distributions}. We note that the  most energetic electrons quickly disappear after the injection is switched off - they cool and accumulate at lower energies. Since the cooling times are much longer for protons than for electrons, the behavior of the proton distribution is quite different. The number of the most energetic protons also decreases for $z > z_{\rm{max}}$, but there are plenty of high-energy protons outside the acceleration region. In one-zone models, where there is no convection term in the transport equation, or it is replaced by an escape term, the particle distributions would be zero in the region where $Q(E,z)\approx0$.

\subsection{Spectral energy distributions}

\begin{figure*}[t]%
\centering
\includegraphics[width=0.49\textwidth, keepaspectratio]{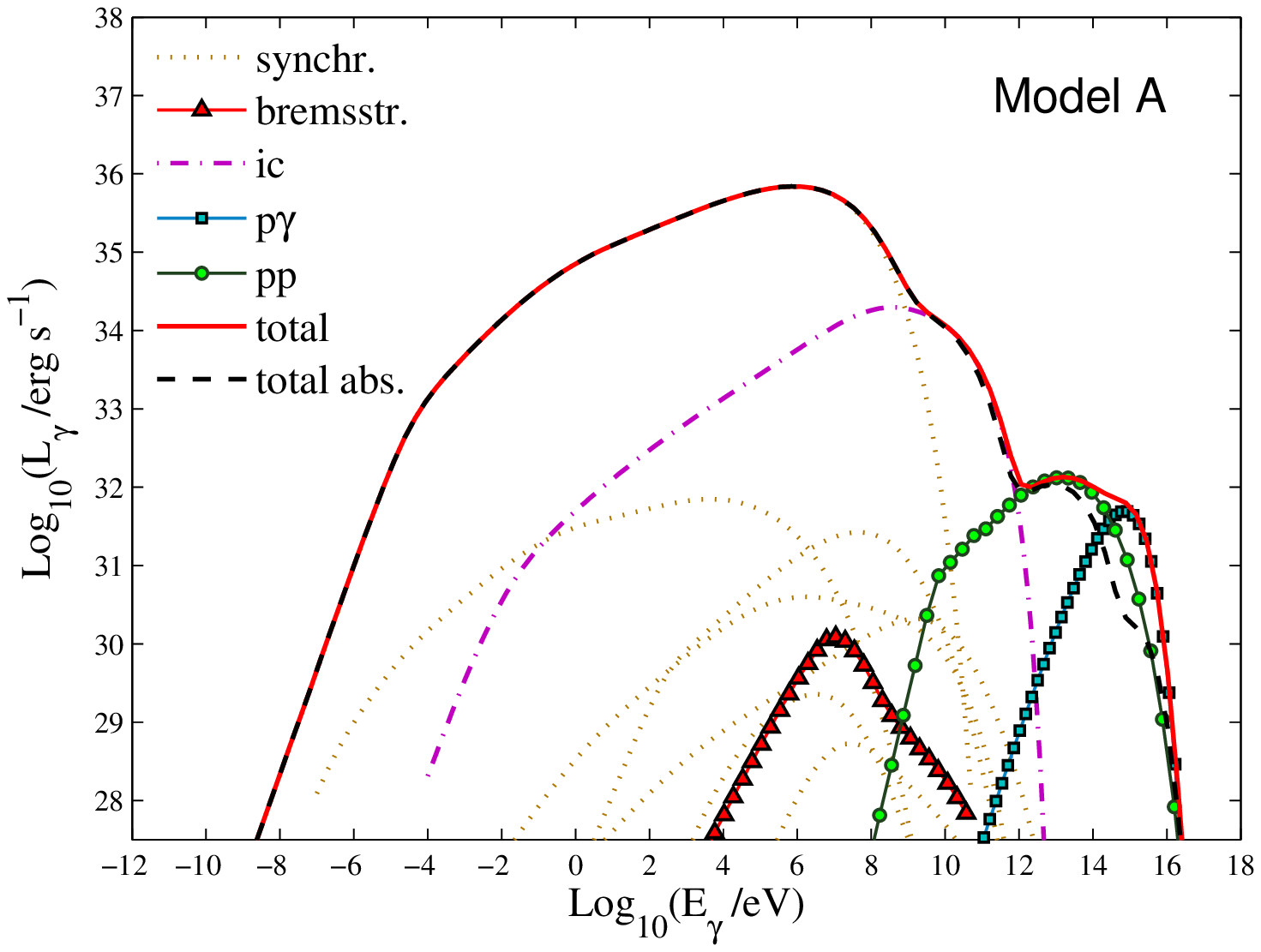}%
\includegraphics[width=0.49\textwidth, keepaspectratio]{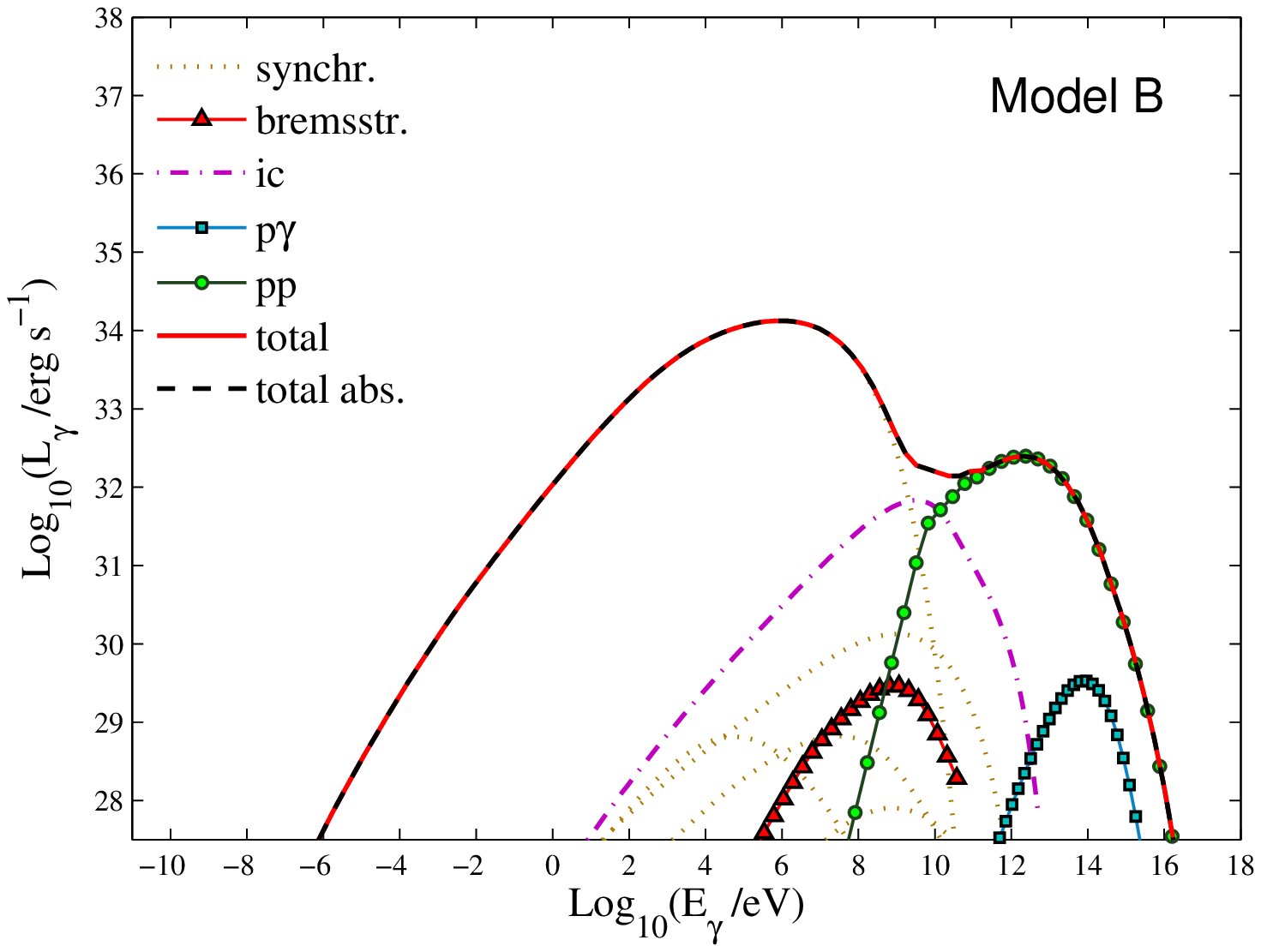}
\includegraphics[width=0.49\textwidth, keepaspectratio]{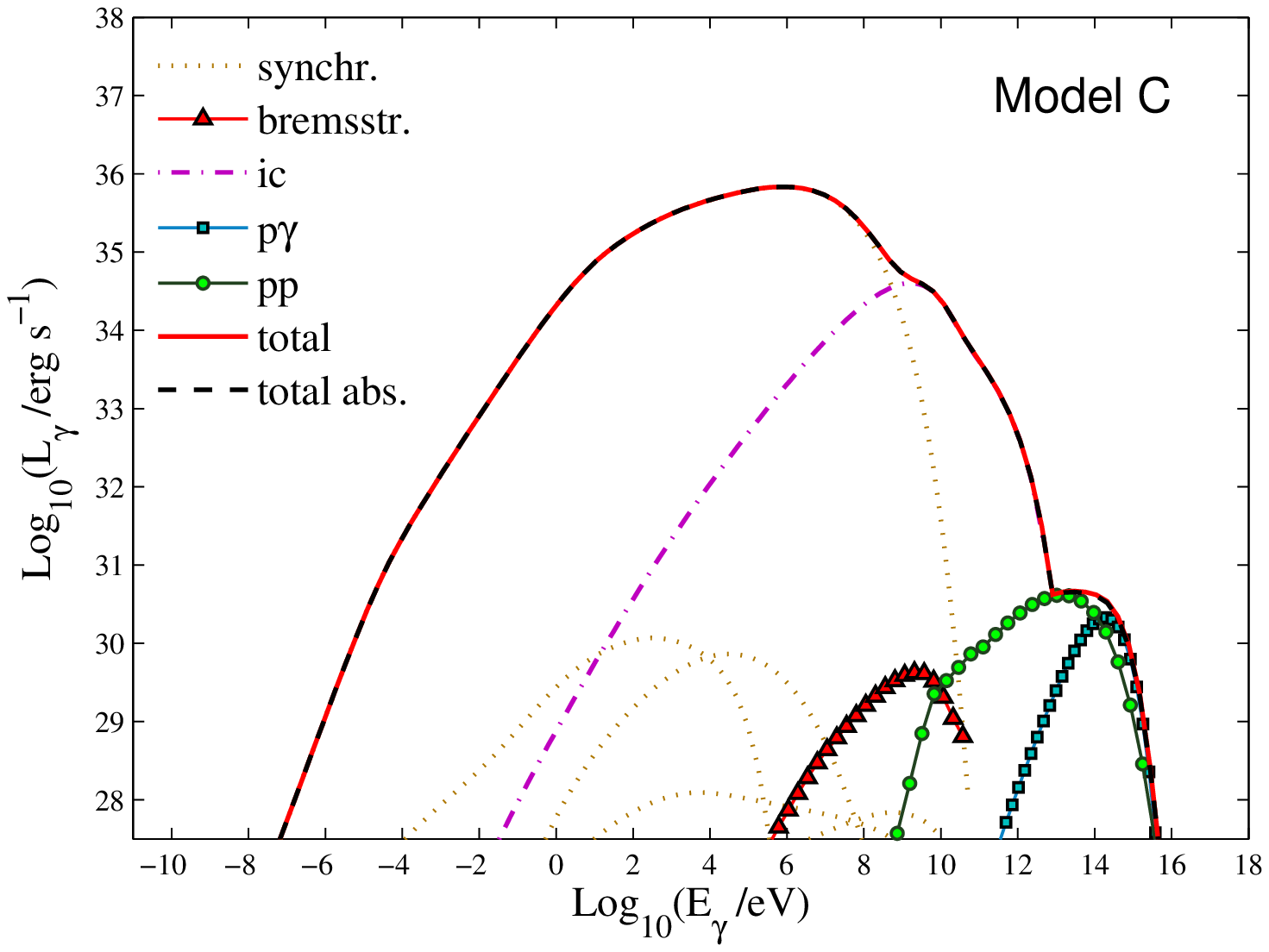}%
\includegraphics[width=0.49\textwidth, keepaspectratio]{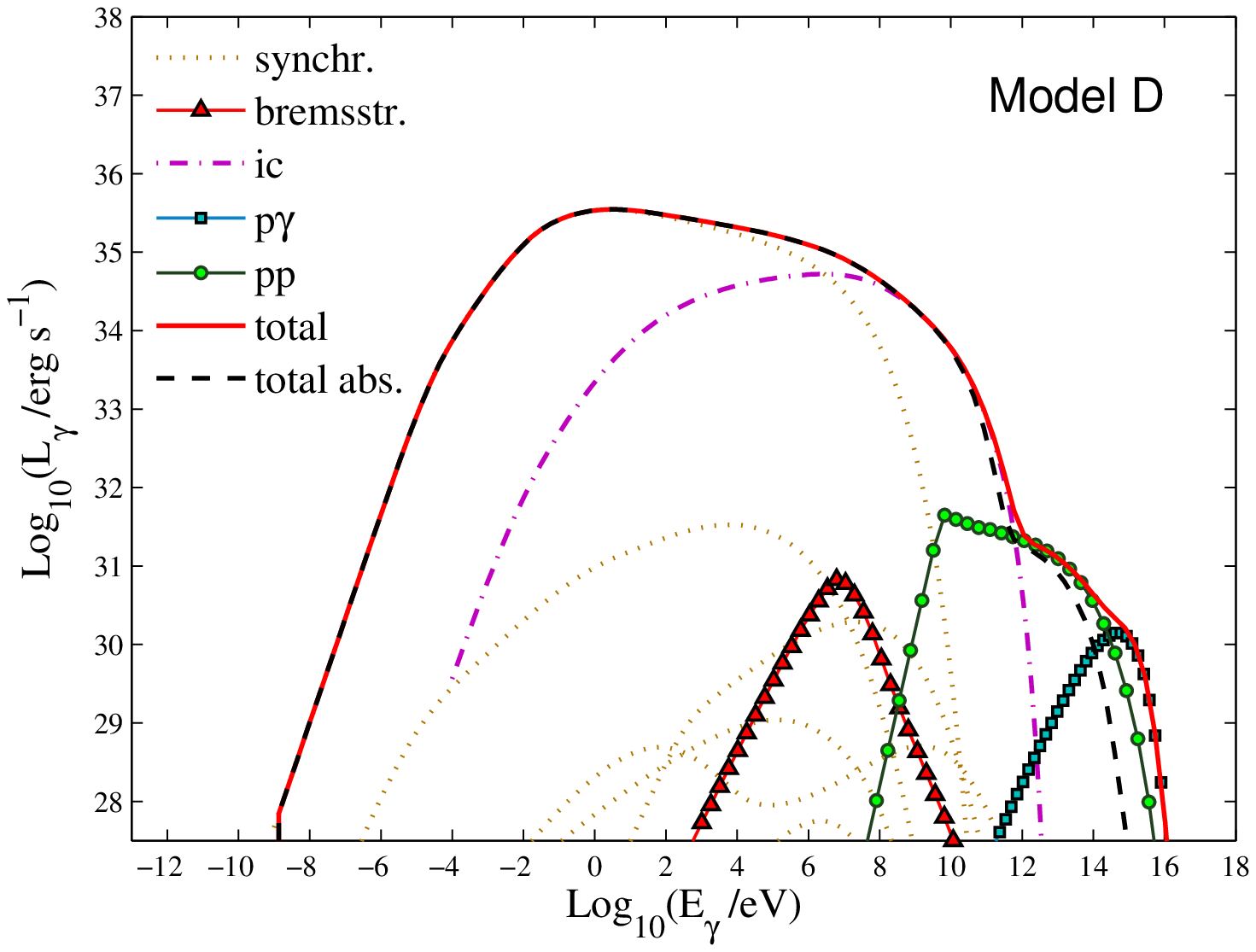}%
\caption{Spectral energy distributions for the four general models of Table \ref{tab:general_models}. The various curves labeled  ``synchr.'' correspond to the synchrotron radiation of primary electrons (the most luminous component), protons, and secondary particles (pions, muons, and electron-positron pairs). Figures are available in color in the electronic version of the manuscript. }%
\label{fig:generalSEDs}%
\end{figure*}

Figure \ref{fig:generalSEDs} shows the SEDs obtained for the four models of Table \ref{tab:general_models}, which have a broad range of spectral shapes. 

In model A, the power injected in relativistic protons is large ($a=1$). The jet  emission up to $\sim 1$ TeV is synchrotron and IC radiation of primary electrons, reaching luminosities of \mbox{$\sim10^{36}$ erg s$^{-1}$} at $\sim10$ MeV. The very high-energy tail of the spectrum is due to proton-proton and proton-photon interactions. 

In model B, most of the energy is transferred to relativistic protons ($a=100$). The synchrotron radiation field of primary electrons is considerably weaker, and so are all other interactions for which this photon field is a target (proton-photon collisions and IC scattering). Furthermore, in this model $m=2$ and the magnetic field strength decreases rapidly with $z$. This also contributes to quenching the synchrotron emissivity. We note that the radiative output of proton-proton collisions is only slightly affected compared to that of model A. 

The only difference between  model C and model A is the extent of the acceleration region. In model C, the base of the region is shifted to slightly higher $z$, and extends up to $10^{11}$ cm. This ``spread'' in the spatial distribution of the relativistic particles affects the proton-proton gamma-ray spectrum. The more extended the acceleration region, the less radiatively efficient this process is. 

Finally, in model D the injection spectral index of the relativistic particles changes from $\alpha = 1.5$ to $\alpha = 2.2$. This turns the electron synchrotron spectrum from hard to relatively soft. The same happens with the gamma-ray emission produced by proton-proton collisions. 

Both Bremsstrahlung and  synchrotron radiation of secondary particles  do not contribute significantly to the SED of any model.

The total luminosity corrected by absorption is also plotted for every model of Fig. \ref{fig:generalSEDs}. The effect of absorption is only noticeable at very high energies in models A and D, where the acceleration region is less extended. 

These four models do not  exhaust the possible parameter space that can be explored. To calculate the SEDs of models A to D, we choose a set of fixed parameters and study the effects of varying the rest of them. Modifying other parameters, such as $q_{\rm{accr}}$ or $\eta$, would introduce further interesting changes in the spectrum, including higher gamma-ray luminosities. This is clearly exemplified by the SEDs calculated to fit the spectrum of XTE J1118+480 presented in Sect. \ref{subsec:XTE_SEDs}. 

\section{The low-mass microquasar XTE J1118+480: characteristic parameters and observations}
\label{sec:XTE}

XTE J1118+480 is an X-ray binary in the Galactic halo. It hosts a low-mass donor star ($M_*\approx0.37M_\odot$) and a black hole of mass $M_{\rm{BH}}\approx8.53 M_\odot$ (Gelino et al. 2006).

The estimated distance to XTE J1118+480 is $d\approx 1.72$ kpc (Gelino et al. 2006). The source is located at high Galactic latitude ($b = +62^\circ$) in a region of low interstellar absorption along the line of sight. This peculiarity of its position allows to perform very clean observations. 

XTE J1118+480 is a transient XRB, spending long periods in quiescence. Since it was first detected in 2000 (Remillard et al. 2000), two outbursts were observed: at the time of its discovery and in 2005. The source was extensively observed at different wavelengths during the two episodes, and on both occasions the spectrum showed the characteristics of the low-hard state. No outflows were directly imaged, but the presence of jets can be inferred from the radio-to-infrared/optical emission (Fender et al. 2001). 

The outburst of 2000 lasted for about seven months. Observational data with very complete spectral coverage were presented and analyzed in Hynes et al. (2000), McClintock et al. (2001), Esin et al. (2001), and Chaty et al. (2003). As shown in Chaty et al. (2003), the SED did not change significantly over a period of about three months.  The second outburst started in January 2005 (Zurita et al. 2005; Pooley 2005; Remillard et al. 2005) and lasted for 1-2 months (Zurita et al. 2006). Radio-to-X-ray data from this epoch were presented in Hynes et al. (2006) and Zurita et al. (2006). 

The radio-to-X-ray spectrum of XTE J1118+480 in outburst has been interpreted as the sum of the black body emission of a thin accretion disk and synchrotron radiation from non-thermal electrons in a jet (see, for example, the works of  Markoff et al. 2001, Yuan et al. 2005, Maitra et al. 2009, and Brocksopp et al. 2010).  Other authors replace the contribution of the jet by that of a hot, optically thin ADAF (see Esin et al. 2001 and Yuan et al. 2005). The ADAF models, however, underpredict the observed radio and UV emission. 

We apply our inhomogeneous jet model to fit the broadband data from the two known outbursts of XTE J1118+480. Differentiating from previous works, we explore the consequences of the injection of non-thermal protons and secondary particles in an extended region of the jet. We also consider the effects of internal and external absorption on the jet emission. 

Table \ref{tab:obs_data} shows a brief log of the observations used for the fits. The data from the 2000 outburst are taken from McClintock et al. (2001), whereas for the 2005 outburst we use the data published in Maitra et al. (2009).\footnote{We extracted the data  with the help of the ADS's Dexter Data Extraction Applet and a script prepared by the authors.} Please refer to these works for details of the instrumental techniques and the reduction process. The UV/X-ray spectrum of the 2000 outburst displays a ``dip'' in the energy range 0.15-2.5 keV. According to Esin et al. (2001), this feature might be caused by absorption in a region of partially ionized gas interposed in the line of sight. Following Maitra et al. (2009), we exclude this energy band from the fit.  

\begin{table}[htb]
%\begin{center}
\caption{Log of the observational data used in the fits.}
\label{tab:obs_data}
\begin{tabular}{@{}l|l|c}
\hline
\hline
&&\\[-0.2cm]
Observation date & Instrument    & Range \\[0.1cm]
\hline
&&\\[-0.2cm]
2000 April 18       & Ryle Telescope 							& 15.2 GHz  \\[0.1cm]
		 								& UKIRT\tablefootmark{a}			& 1-5 $\mu$m \\[0.1cm]
		 								& HST\tablefootmark{b}	  		& 1155-10250 \r{A}\\[0.1cm]	
		 								& EUVE\tablefootmark{c}	  		& 0.1-0.17 keV \\[0.1cm]
		 								& \emph{Chandra}							& 0.24-7 keV \\[0.1cm]
		 								& RXTE - PCA\tablefootmark{d}	& 2.5-25 keV \\[0.1cm] 
		 								& RXTE - HEXTE\tablefootmark{e}	& 15-200 keV \\[0.1cm] 
\hline
&&\\[-0.2cm]
2005 January 23 & Ryle Telescope 			& 15.2 GHz  \\[0.1cm]
		 						& UKIRT								& J, H, K-band \\[0.1cm]
		 						& Liverpool Telescope	& V, B-band \\[0.1cm]	
		 						& RXTE 								& 3-70 keV \\[0.1cm]  
\hline
\hline
\end{tabular}
\tablefoot{
\tablefoottext{a}{United Kingdom Infrared Telescope}
\tablefoottext{b}{Hubble Space Telescope}
\tablefoottext{c}{Extreme Ultraviolet Explorer}
\tablefoottext{c}{Rossi X-Ray Timing Explorer - Proportional Counter Array}
\tablefoottext{c}{Rossi X-Ray Timing Explorer - High Energy X-ray Timing Experiment}
}
%\end{center}
\end{table}

%______________________________________________________________

\section{Fits to the SED of XTE J1118+480 in low-hard state }

\subsection{Accretion disk model}

The shape of the observed SED  of XTE J1118+480 in the IR-optical/UV band strongly suggests that it is produced by thermal radiation from an accretion disk; we use a simple model to fit this component.  We parameterize the radial dependence of the temperature as (e.g. Frank et al. 2002)

\begin{equation}
T(R) = \frac{T_{\rm{{max}}}}{0.488}\left(\frac{R}{R_{\rm{{in}}}}\right)^{-3/4}\left(1-\sqrt{\frac{R_{\rm{{in}}}}{R}}\,\right)^{1/4},
	\label{eq:disc_temperature_profile}
\end{equation}

\vspace{0.2cm}

\noindent where $R_{\rm{{in}}}$ is the disk inner radius, and $T_{\rm{{max}}}$ is the maximum disk temperature reached at $R_{\rm{{max}}}=(49/36)R_{\rm{{in}}}$. This temperature profile is consistent with that of a standard optically thick, geometrically thin accretion disk (Shakura \& Sunyaev 1973).

Every annulus of the disk radiates as a black body at the local temperature $T(R)$. The observed flux  at energy $E_\gamma$ is then

\begin{equation}
F_{\rm{d}}(E_\gamma) = 2\pi\frac{\cos\theta_{\rm{d}}}{d^2}\int_{R_{\rm{{in}}}}^{R_{\rm{{out}}}} B(E_\gamma,R)\,R\,dR,
	\label{eq:disc_flux}
\end{equation}

\vspace{0.2cm}

\noindent where $B(E_\gamma,R)$ is the Planck function

\begin{equation}
B(E_\gamma,R) = \frac{2}{c^2h^3}\frac{E_\gamma^3}{\exp\left[E_\gamma/kT(R)\right]-1}
	\label{eq:planck_function}
\end{equation}

\vspace{0.2cm}

\noindent and $\theta_{\rm{d}}$ is the inclination angle of the disk with respect to the line of sight. We assume that the disk is perpendicular to the jet, so $\theta_{\rm{d}}=\theta_{\rm{jet}}$. 

The power emitted per unit  area of the disk is \mbox{$D(R)=\sigma_{\rm{SB}}T(R)^4$}, where $\sigma_{\rm{SB}}$ is the Stefan-Boltzmann constant. The total luminosity of the disk can be calculated by integrating $D(R)$ over the two faces of the disk

\begin{equation}
L_{\rm{d}} = 2\times 2\pi\sigma_{\rm{SB}}\int_{R_{\rm{{in}}}}^{R_{\rm{{out}}}} T(R)^4\,R\,dR \approx \frac{4\pi}{3}\sigma_{\rm{SB}}\left(\frac{T_{\rm{{max}}}}{0.488}\right)^4R_{\rm{{in}}}^2.
	\label{eq:disc_luminosity_from_temperature}
\end{equation}

\vspace{0.2cm}

\noindent The approximation is valid for $R_{\rm{{out}}}>>R_{\rm{{in}}}$.

In steady state, half of the gravitational energy lost by the infalling matter is radiated in the disk. Then 

\begin{equation}
L_{\rm{d}} = \frac{1}{2}\frac{GM_{\rm{{BH}}}\dot{M}}{R_{\rm{{in}}}} = \frac{1}{2}\frac{R_{\rm{{grav}}}}{R_{\rm{{in}}}}\dot{M}c^2.
	\label{eq:disc_luminosity_from_grav_energy}
\end{equation}

\vspace{0.2cm}

\noindent Eqs. (\ref{eq:disc_luminosity_from_temperature}) and (\ref{eq:disc_luminosity_from_grav_energy}) provide an estimation of the accretion power

\begin{equation}
\dot{M}c^2 = \frac{8\pi}{3}\sigma_{\rm{SB}}\left(\frac{T_{\rm{{max}}}}{0.488}\right)^4\frac{R_{\rm{{in}}}^3}{R_{\rm{{grav}}}}.
	\label{eq:accretion_power_from_disc_luminosity}
\end{equation}

\noindent The photons from the disk irradiate the jet. Since the disk is considerably luminous, the interaction of the disk photons with the relativistic particles in the jet cannot be neglected. We consider the IC scattering of the disk photons by the primary electrons and muons in the jet. In the observer frame, the interaction is anisotropic, and the dependence of the Klein-Nishina cross-section on the collision angle must be considered. We calculate the anisotropic IC spectrum as in Dermer \& Schlickeiser (1993). To  simplify the formulas, the disk radiation field at fixed radius is approximated as a monoenergetic field at energy $\left\langle \epsilon \right\rangle = 2.7kT(R)$, where $k$ is the Boltzmann constant. Under the same approximations, an analytical expression for the energy-loss rate for this process is obtained in B\"ottcher et al. (1997).\footnote{Eqs. (15) and (16) of B\"ottcher et al. (1997) fail for values of their parameter $a$ very close to unity. Since this interval does not add significantly to the value of the cooling rate, we disregard it in the calculation.}

High-energy photons emitted in the jet can be absorbed by the accretion disk radiation field to create electron-positron pairs. As IC scattering, this process is also highly anisotropic. We estimate the opacity for photons emitted along the jet axis as a function of energy and $z$ following Becker \& Kafatos (1995). For this calculation, we also approximate the disk photon field at fixed radius as monoenergetic. 

\subsection{Parameters of the fit}

In Table \ref{tab:fixed-parameters}, we list the values of the model parameters that we keep fixed during the fit. Some of them, such as the black hole mass and the distance to Earth, are inferred from observations; we take their values from the literature about the source. For the remaining parameters, we have no information about XTE J1118+480 (such as the bulk Lorentz factor), or they are specific to our model. For them, we adopt typical values for other microquasars, or simply estimates.  

The parameters in Table \ref{tab:free-parameters} were varied to obtain the best-fit models. The inner radius $R_{\rm{in}}$ and the temperature $T_{\rm{max}}$ determine the spectrum of the accretion disk. The value of $q_{\rm{{accr}}}$ follows from Eqs. (\ref{eq:accretion_power}) and (\ref{eq:accretion_power_from_disc_luminosity}).  The rest of the parameters of the jet model (see Table \ref{tab:dependent-parameters}) are calculated from the fixed and free parameters using the equations of Sect. \ref{subsec:jet_parameters}. 

\begin{table}[htb]
%\begin{center}
\caption{Values of the fixed parameters of the model.}
\label{tab:fixed-parameters}
\begin{tabular}{@{}p{4.3cm}|c|c}
\hline
\hline
&&\\[-0.2cm]
Parameter & Symbol & Value\\
&&\\[-0.3cm]
\hline 
&&\\[-0.2cm]
Distance to Earth & $d$ & 1.72 kpc\tablefootmark{a} \\ [0.1cm]
Black hole mass & $M_{\rm{BH}}$ & $8.5 M_\odot$\tablefootmark{a}  \\[0.1cm]
Disk outer radius & $R_{\rm{out}}$ & $7\times 10^4 R_{\rm{grav}}$\tablefootmark{b}\\ [0.1cm]
Jet viewing angle  & $\theta_{\rm{jet}}$ & 30$^\circ$ \\[0.1cm]
Jet bulk Lorentz factor  & $\Gamma_{\rm{jet}}$  & 2  \\[0.1cm]
Jet injection distance & $z_0$ & $50 R_{\rm{grav}}$  \\[0.1cm]
Jet termination distance & $z_{\rm{end}}$ & $10^{12}$ cm  \\[0.1cm]
Jet initial radius & $r_0$  & $0.1z_0$  \\  [0.1cm]
Ratio $L_{\rm{rel}}/L_{\rm{jet}}$ & $q_{\rm{rel}}$ & 0.1  \\[0.1cm]
Magnetic field decay index & $m$ & 1.5  \\[0.1cm]
Particle injection spectral index  & $\alpha$ & $1.5$  \\[0.1cm]
Acceleration efficiency & $\eta$ & $0.1$  \\[0.1cm]
\hline
\hline
\end{tabular}
\tablefoot{
\tablefoottext{a}{Gelino et al. (2006).}
\tablefoottext{b}{Chaty et al. (2003).}
}
%\end{center}
\end{table}

\begin{table}[htb]
%\begin{center}
\caption{Values of the free parameters for the best-fit models.}
\label{tab:free-parameters}
\begin{tabular}{@{}p{3.3cm}|c|c|c}
\hline
\hline
%Free parameters & & \\[0.1cm]
Parameter & Symbol &  \multicolumn{2}{c}{Value}\\
\hline 
&&\\[-0.3cm] 
& & 2000 & 2005 \\
\hline
&&\\[-0.2cm] 
Disk inner radius & $R_{\rm{in}}$ & $164\,R_{\rm{grav}}$ & $44\,R_{\rm{grav}}$\\ [0.1cm]  
Disk maximum temperature & $T_{\rm{max}}$ & 22.4 eV & 46.5 eV\\[0.1cm]  
Ratio $L_{\rm{jet}}/L_{\rm{accr}}$ & $q_{\rm{jet}}$ & 0.16 & 0.16\\[0.1cm]
Ratio $U_B/U_{\rm{k}}$ at $z_{\rm{acc}}$ & $\rho$ & 0.5  & 0.85\\[0.1cm]
Ratio $L_p/L_e$  & $a$ & 12.2 & 25.5 \\[0.1cm]
End of acceleration region & $z_{\rm{end}}$ &  $8.2\times 10^{9}$ cm & $1.5\times 10^{10}$ cm\\[0.1cm]  
Minimum energy primary protons and electrons & $E_{\rm{min}}$ & $86\,m_{(p,e)}c^2$ & $150\,m_{(p,e)}c^2$ \\[0.2cm]
\hline
\hline
\end{tabular}
%\end{center}
\end{table}

\begin{table}[htb]
%\begin{center}
\caption{Best-fit model parameters calculated from the free parameters.}
\label{tab:dependent-parameters}
\begin{tabular}{@{}p{2.25cm}|c|c|c}
\hline
\hline
Parameter & Symbol & \multicolumn{2}{c}{Value}\\
\hline
&&\\[-0.3cm] 
 & & 2000 & 2005 \\
\hline
&&\\[-0.2cm] 
Accretion power & $L_{\rm{accr}}$ &  $2.6\times10^{38}$ erg s$^{-1}$ & $9.9\times10^{37}$ erg s$^{-1}$\\[0.1cm]
Jet power & $L_{\rm{jet}}$ &  $2.1\times10^{37}$ erg s$^{-1}$ & $8.1\times10^{36}$ erg s$^{-1}$\\[0.1cm]
Power relativistic protons  & $L_p$ &  $1.9\times10^{36}$ erg s$^{-1}$ & $7.8\times10^{35}$ erg s$^{-1}$\\[0.1cm]
Power relativistic electrons  & $L_e$ &  $1.6\times10^{35}$ erg s$^{-1}$ & $3.1\times10^{34}$ erg s$^{-1}$\\[0.1cm]
Magnetic field jet base & $B_0$ &  $1.3\times10^7$ G & $7.9\times10^6$ G\\[0.1cm]
Base of the acceleration region & $z_{\rm{acc}}$ &  $2.8\times10^8$ cm & $1.6\times10^8$ cm\tabularnewline[0.1cm]    
\hline
\hline
\end{tabular}
%\end{center}
\end{table}

%______________________________________________________________

\subsection{Best-fit spectral energy distributions}
\label{subsec:XTE_SEDs}

\begin{figure*}[htbp]%
\centering
\includegraphics[width=0.48\textwidth]{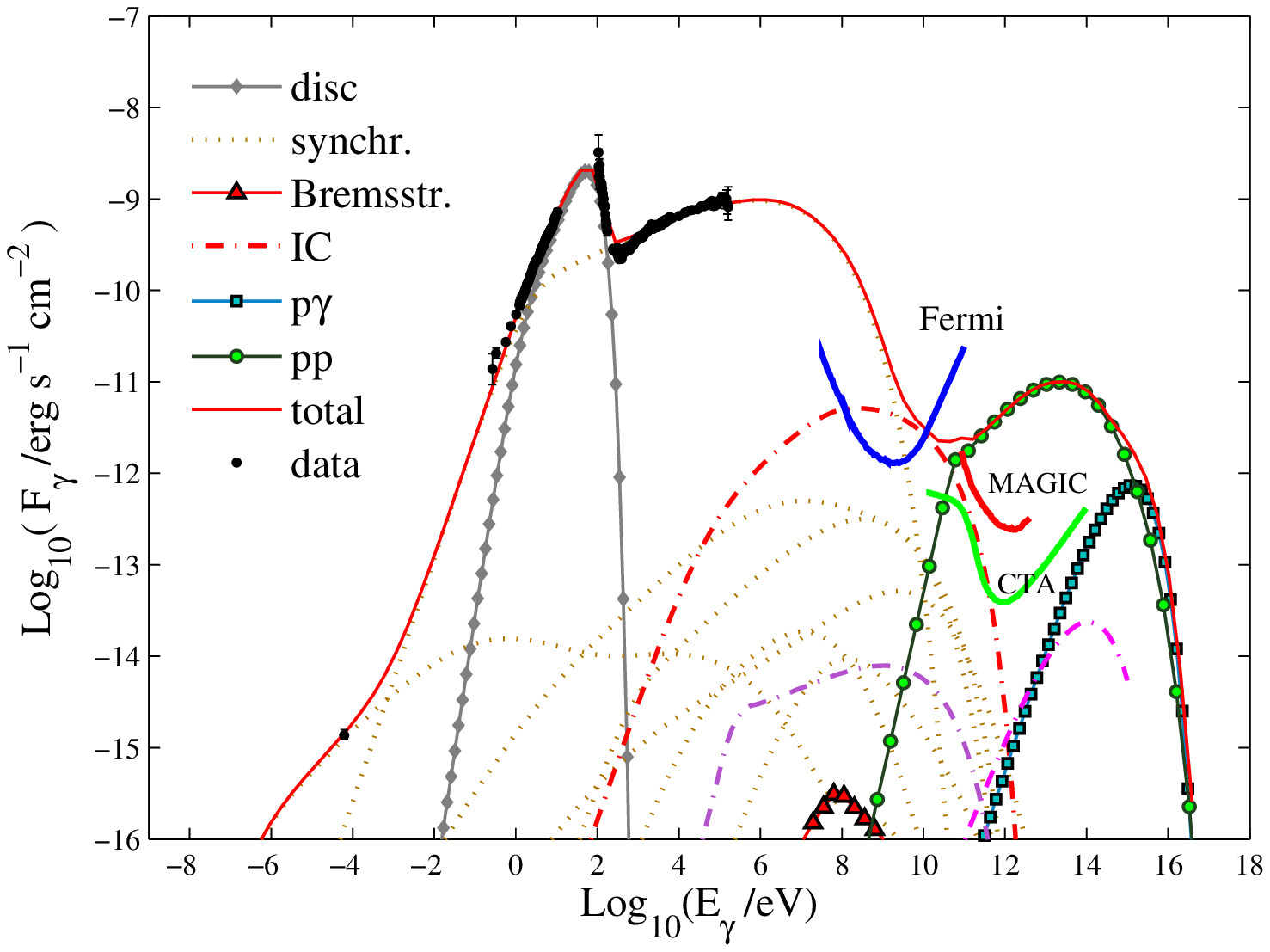}%
\includegraphics[width=0.48\textwidth]{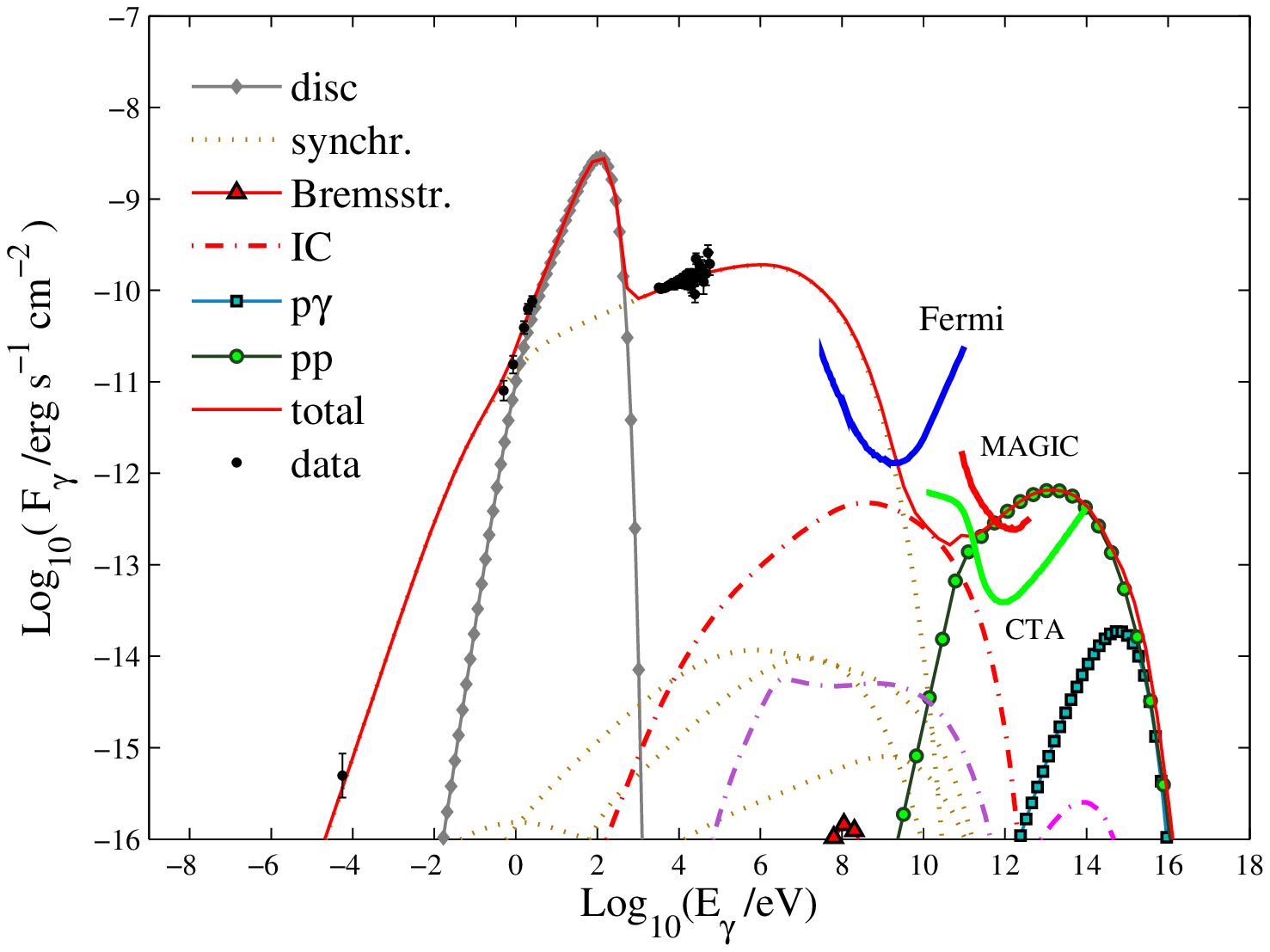}%
\caption{ Best-fit SEDs for the outbursts of 2000 (left) and 2005 (right) of XTE J1118+480. The  various dotted curves labeled under the legend ``synchr.'' correspond to the synchrotron radiation of primary electrons (the most luminous component), protons, and secondary particles (pions, muons, and electron-positron pairs). The dash-dotted curves labeled ``IC'' correspond to the SSC luminosity of primary electrons (red), and the external IC luminosities of primary electrons (violet) and muons (magenta). We also show the sensitivity curves of Fermi-LAT (1 yr exposure,  5$\sigma$), MAGIC II (50 h exposure), and CTA (50 h exposure). Figures are available in color in the electronic version of the manuscript. }%
\label{fig:SEDsXTE}%
\end{figure*}

Figure \ref{fig:SEDsXTE} shows the best-fit SEDs for the 2000 and 2005 outbursts of XTE J1118+480. We obtain $\chi_\nu^2 = 1.99$ and $\chi_\nu^2 = 0.56$ for the chi-squared per degree of freedom, respectively. The best-fit parameters are listed in Tables \ref{tab:fixed-parameters}-\ref{tab:dependent-parameters}. 

\begin{figure*}[t]%
\centering
\includegraphics[width=0.49\textwidth, keepaspectratio, trim = 0 0 0 0, clip]{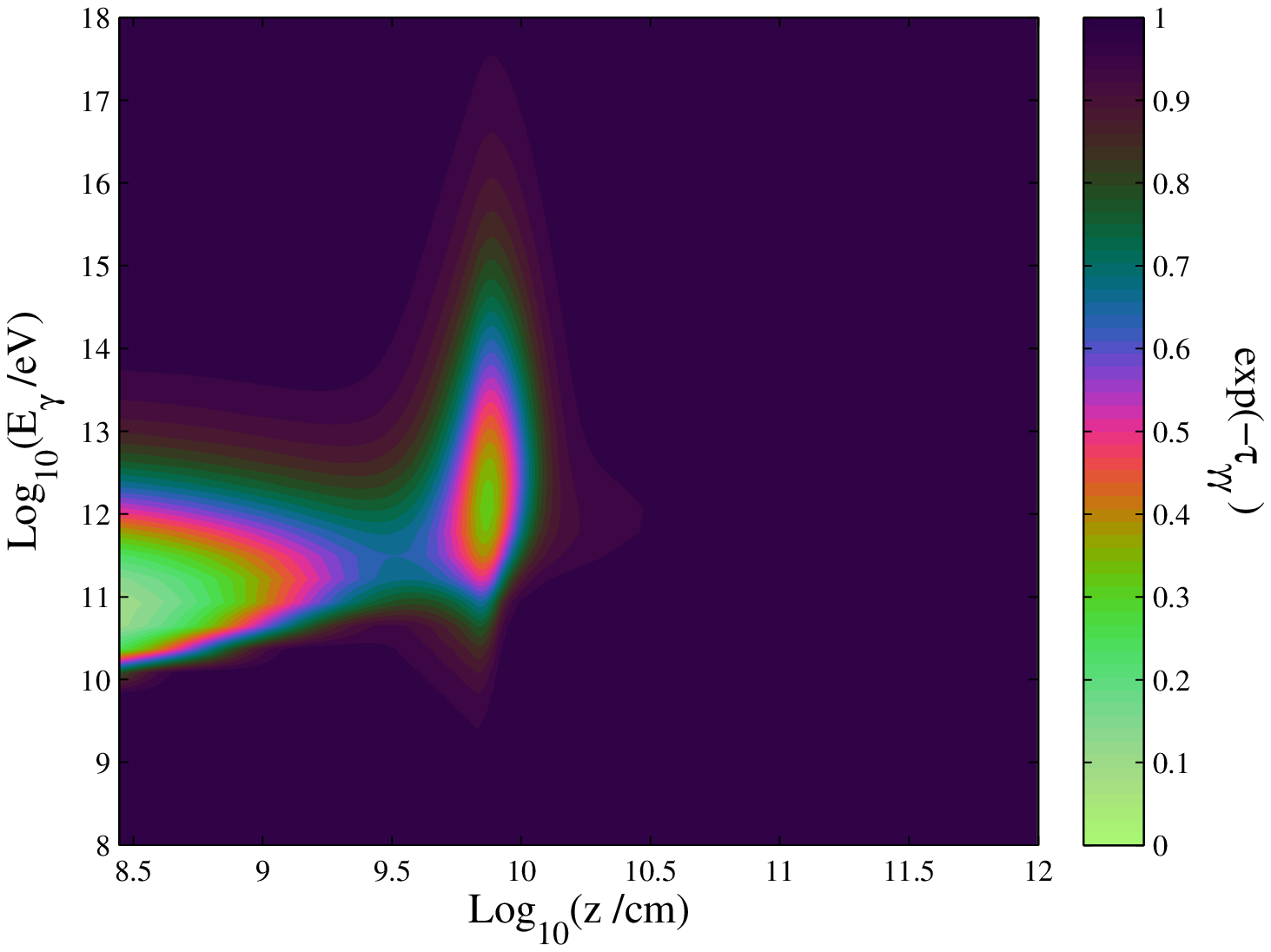}%
\includegraphics[width=0.49\textwidth, keepaspectratio, trim = 0 0 0 0, clip]{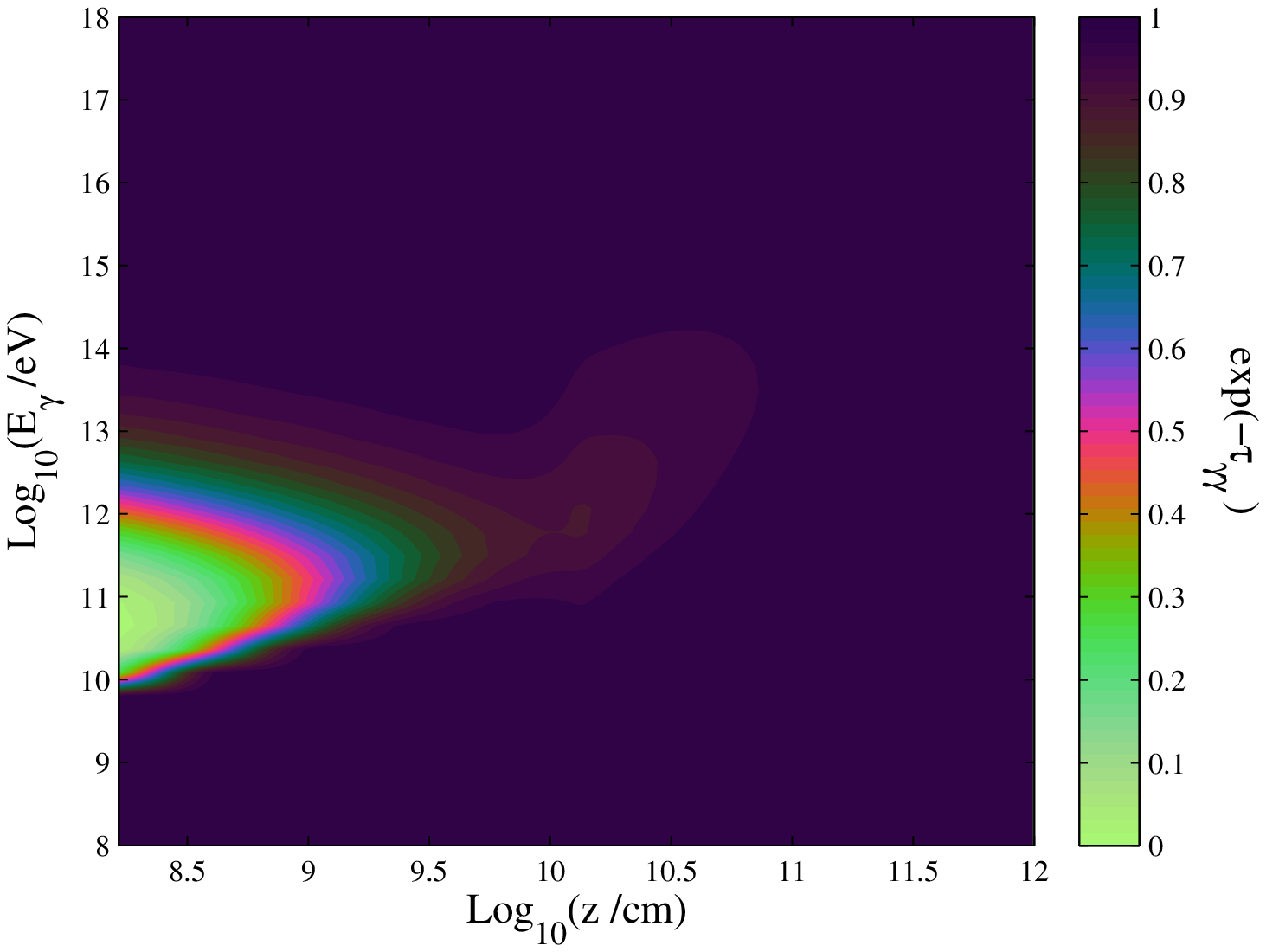}%
\caption{Total attenuation factor $\exp({-\tau_{\gamma\gamma}})$ caused by photon-photon annihilation for the best-fit SEDs of the 2000 (left) and 2005 (right) outbursts of XTE J1118+480. Figure available in color in the electronic version of the manuscript.}%
\label{fig:tau_XTE}%
\end{figure*}

The value of the maximum temperature of the disk agrees with previous works (see for example McClintock et al. 2001, Markoff et al. 2001, Chaty et al. 2003, and Maitra et al. 2009). In the case of the 2000 outburst, $T_{\rm{max}}$ is tightly constrained by the data, which clearly indicate the position of the peak of the multicolor black-body component. The value of the inner radius of the disk is not well-constrained by the observations, and depends on other details of the model as we discuss below. There is no agreement in the value of this parameter reported in the literature. 

We emphasize that our disk model is simple. We do not include effects such as irradiation of the outer disk, or the transition to an ADAF in the surroundings of the black hole. We do not attempt, therefore, to tightly constrain any characteristic parameter of the disk, but just to account roughly for the thermal component observed in the SED. 

The typical accretion power of XRBs in low-hard state is $L_{\rm{accr}}\approx 0.01-0.1L_{\rm{Edd}}$. The value of $q_{\rm{accr}}\approx 0.08$ that  we obtain from for the 2005 outburst is within this range. It is higher for the 2000 outburst, $q_{\rm{accr}}\approx 0.22$, mainly because the  disk inner radius is larger.  Once $T_{\rm{max}}$ and  $\theta_{\rm{d}}$ are fixed, $R_{\rm{in}}$ determines the normalization of the spectrum. The maximum temperature of the disk for the 2000 outburst is well-constrained; the inclination angle, however, is unknown.\footnote{There are estimates of the orbital inclination for XTE J1118+480, e.g. Gelino et al (2006). However, there is no compelling reason to assume that the disk lies in the orbital plane of the binary. See, for example, Maccarone (2002), Butt et al. (2003), and Romero \& Orellana (2005) for a discussion of misaligned microquasars. } The best-fit value of $R_{\rm{in}}$ then depends on the value chosen for $\theta_{\rm{d}}$, which we assumed to equal $30^\circ$. Given the strong dependence of the accretion power on $R_{\rm{in}}$ (see Eq. \ref{eq:accretion_power_from_disc_luminosity}), small variations in this parameter yield significant changes in $q_{\rm{accr}}$.

According to our modeling, there are no great differences in the physical conditions of the jets during the two outbursts. The radio and X-ray emission is fitted by the synchrotron spectrum of primary electrons, plus some contribution at low energies of secondary pairs created by $\gamma\gamma$ annihilation.
The IR-optical-UV range has significant contribution from the accretion disk.  The synchrotron emission of secondary particles is negligible (except in the case mentioned above), as well as Bremsstrahlung radiation of primary electrons. The IC scattering off the jet photon field by primary electrons contributes in a narrow energy range at $\sim10$ GeV in the case of the 2000 outburst. The SED above \mbox{$\sim1$ GeV} is completely dominated by gamma rays from the decay of neutral pions created in proton-proton collisions.

The attenuation factor $e^{-\tau_{\gamma\gamma}}$ as a function of energy and $z$ is plotted in Fig. \ref{fig:tau_XTE}. The main source of absorbing photons is the accretion disk; the internal radiation field of the jet only adds a ``bump'' at high energies. The optical depth is large only near the base of the acceleration region. Gamma rays with energies \mbox{10 GeV $\la E_\gamma \la$ 1 TeV } are mostly absorbed in this zone. The development of electromagnetic cascades, however, is suppressed by the strong magnetic field (see Pellizza et al. 2010). 

The total luminosities are unmodified by absorption. As discussed in Sect. \ref{sub-sec:particle_distributions}, there are many high-energy protons that produce gamma rays through proton-proton collisions outside the acceleration region. This radiation escapes unabsorbed because the density of disk photons is low at high $z$. 

The sensitivity curves of the Large Area Telescope (LAT) on board the Fermi satellite, the Major Atmospheric Gamma-ray Imaging Cherenkov Telescope II (MAGIC II), and the predicted for the future Cherenkov Telescope Array (CTA) are also plotted in Fig. \ref{fig:SEDsXTE}. According to our results, a future outburst of the source with emission levels comparable to those of 2000 and 2005, would be detectable in gamma rays by ground-based observatories such as MAGIC II and CTA.  

To date, no low-mass X-ray binaries have been detected with Cherenkov telescopes.  Their observation is difficult because in general they are transient sources. Negative detections of four low-mass microquasars with HESS were reported in Chadwick et al. (2005). Three of them, however, were in the high-soft state; there are no available simultaneous X-ray data for the fourth source, GX 339-4, but it was apparently in a low-luminosity state. 

In the context of the model presented here, observations of low-mass microquasars at very-high energies would help us to constrain the hadronic content of the jets, since above \mbox{$\sim100$ GeV} the predicted emission is completely generated by proton-proton interactions.

In addition, no low-mass X-ray binaries have been detected at high-energies. There is one Fermi gamma-ray source, 1FGL J1227.9-4852, that might be the counterpart to the bright low-mass X-ray binary XSS J12270-4859, but the association remains unclear (Falanga et al. 2010, Hill et al. 2011).

The detectability in the Fermi-LAT band basically depends on the position of the synchrotron cutoff. We fix $\eta=0.1$ for the acceleration efficiency, which yields a high maximum energy for the electrons, and the synchrotron emission extends into the MeV energy range. Observations with Fermi, then, could help us to investigate the efficiency of particle acceleration in jets of microquasars. 

%______________________________________________________________

\section{Concluding remarks}

We have developed an inhomogeneous lepto-hadronic jet model, and presented some general results for the relativistic particle distributions, cooling rates, and the jet radiative spectrum. The model can satisfactorily reproduce the observed SED of the low-mass microquasar XTE J1118+480 in low-hard state. The best-fit SEDs have revealed some interesting consequences of the injection of relativistic protons in an inhomogeneous extended region. In particular, we have demonstrated that the gamma-ray emission can escape absorption, and that the source might be detectable when a new outburst of similar characteristics occurs. 

With some improvements, the same jet model can be applied to active galactic nuclei and gamma-ray bursts. The main modifications concern the calculation of the particle distributions. Eq. (\ref{eq:transport_equation}) would have to be modified to a covariant version, which could then be applied to outflows with very high bulk Lorentz factors. In addition, within environments where the radiation field is ultra-dense (as for the jets of long gamma-ray bursts), the coupling among the kinetic equations for particles and photons could not be ignored. A further generalization of the transport equation would be to include time dependence, allowing the study of flares. Other interesting improvements to the model would be in the treatment of the jet kinematics and energetics, to develop a jet model that consistently couples the dynamical and radiative aspects. All these topics are part of our research project, and will be addressed in future works.

%______________________________________________________________

\begin{acknowledgements}

G.S.V. thanks Iva \v{S}nidari\'c  for the data on the sensitivity of MAGIC II. This work was supported by CONICET grant PIP 0078, by and the Spanish Ministerio de Ciencia e Innovaci\'on (MINCINN) under grant AYA 2010-21782-C03-01.

\end{acknowledgements}

%______________________________________________________________

\end{document}